\documentclass[letterpaper,aps,showpacs,showkeys,eqsecnum]{revtex4}
\usepackage{anysize}
\marginsize{2cm}{2cm}{1cm}{1cm}
\usepackage{bm}
\usepackage{amsfonts}
\usepackage{amsmath}
\usepackage{amssymb}
\usepackage{graphicx}
\usepackage{feynmp}
\usepackage[usenames,dvipsnames]{xcolor}
\usepackage[colorlinks=true,linkcolor=Blue,citecolor=Blue,linktocpage,urlcolor=Blue]{hyperref}
\newcommand{\aeq}{\begin{equation}}
\newcommand{\ceq}{\end{equation}}
\newcommand{\aec}{\begin{eqnarray}}
\newcommand{\cec}{\end{eqnarray}}
\newcommand{\ase}{\begin{subequations}}
\newcommand{\cse}{\end{subequations}}

\renewcommand{\(}{\left(}
\renewcommand{\)}{\right)}
\renewcommand{\[}{\left[}
\renewcommand{\]}{\right]}

\renewcommand{\a}{\alpha}
\renewcommand{\b}{\beta}

\newcommand{\m}{\mu}
\newcommand{\n}{\nu}
\renewcommand{\o}{\omega}
\newcommand{\g}{\gamma}
\renewcommand{\d}{\delta}
\newcommand{\h}{\eta}
\newcommand{\z}{\zeta}
\newcommand{\f}{\phi}

\newcommand{\y}{\psi}
\renewcommand{\l}{\lambda}
\newcommand{\s}{\sigma}
\renewcommand{\r}{\rho}

\newcommand{\q}{\theta}
\newcommand{\e}{\epsilon}

\renewcommand{\t}{\tau}
\newcommand{\p}{\pi}

\newcommand{\G}{\Gamma}

\newcommand{\D}{\Delta}

\newcommand{\pd}{\partial}

\DeclareFontFamily{OT1}{pzc}{}
\DeclareFontShape{OT1}{pzc}{m}{it}{<-> s * [1.10] pzcmi7t}{}
\DeclareMathAlphabet{\mathpzc}{OT1}{pzc}{m}{it}
\begin{document}
\title{Status of the lower spins in the   Rarita-Schwinger
four-vector spinor $\psi_\mu$ within the method of the
combined Lorentz- and Poincar\'e invariant projectors}

\author{E. G. Delgado-Acosta}\email{german@ifisica.uaslp.mx}
\affiliation{Instituto de F\'{\i}sica, Universidad Aut\'onoma de San Luis
Potos\'{\i},
Av. Manuel Nava 6, San Luis Potos\'{\i}, S.L.P. 78290, M\'exico}

\author{V. M. Banda-Guzm\'an}\email{vmbg@dec1.ifisica.uaslp.mx}
\affiliation{Instituto de F\'{\i}sica, Universidad Aut\'onoma de San Luis
Potos\'{\i},
Av. Manuel Nava 6, San Luis Potos\'{\i}, S.L.P. 78290, M\'exico}

\author{M. Kirchbach}\email{mariana@ifisica.uaslp.mx}
\affiliation{Instituto de F\'{\i}sica, Universidad Aut\'onoma de San Luis
Potos\'{\i},
Av. Manuel Nava 6, San Luis Potos\'{\i}, S.L.P. 78290, M\'exico}

\date{\today}

\begin{abstract}
We investigate the status of the lower spin-$1/2$ companions to spin-$3/2$ within the four-vector 
spinor, $\psi_\mu$. According to its reducibility, 
$\psi_\mu\longrightarrow \left[(1/2,1)\oplus (1,1/2)\right]\oplus [(1/2,0)\oplus (0,1/2)]$
this representation space contains two spin-$1/2$ sectors,  the first one transforming as a genuine 
Dirac-spinor, $(1/2,0)\oplus (0,1/2)$, and the second as the companion to spin-$3/2$ in
$(1/2,1)\oplus (1,1/2)$. In order to correctly  identify the covariant spin-$1/2$ degrees of freedom 
in the Rarita-Schwinger field of interest we exploit the properties of the Casimir invariants of the 
Lorentz algebra to distinguish between the irreducible Dirac- and $(1/2,1)\oplus (1,1/2)$ representation 
spaces and construct corresponding momentum-independent (static) projectors which we then combine with a 
dynamical spin-$1/2$ Poincar\'e covariant projector, based on the two Casimir invariants of the Poincar\'e 
algebra- the squared momentum, and the squared Pauli-Lubanski vector. In so doing we obtain two spin-$1/2$ 
wave equation, and prove them to describe causal propagation of the wave fronts within an electromagnetic 
field. We furthermore calculate Compton scattering off each one of the above states, and find that the 
amplitudes corresponding to the first spin-$1/2$ are identical to those of a Dirac particle and conclude 
on the observability of this state. Also for the second spin-$1/2$ we find finite cross sections in all 
directions in the ultrarelativistic limit, and conclude that its observability is not excluded neither by 
causality of propagation within an electromagnetic environment, nor by  unitarity of the Compton scattering 
amplitudes in the ultraviolet. Finally, we  notice that  the method of the combined   Lorentz- and Poincar\'e
invariant projectors could be instrumental in opening  a new avenue  toward the consistent description of 
any spin by means of second order Lagrangians written in terms of sufficiently large reducible representation 
spaces equipped with separate Lorentz-- and Dirac indices. Specifically, the antisymmetric Lorentz tensor of 
second rank with either Dirac-spinor components, $\psi_{\left[ \mu\nu \right]}$, or, with four-vector 
spinor components, $\psi_{\left[ \mu \nu \right]\eta }$, can be employed in the description of single 
spin $(3/2,0)\oplus (0,3/2)$, or spin-$5/2$ as part of $(2,1/2)\oplus (1/2,2)$, respectively.

%In a similar way, spin-$5/2$ can
%be seen as part of
%the totally antisymmetric tensor of second rank with four-vector spinor
%components,
%$\psi_{\left[\mu \nu \right]\eta }$ and specifically as a resident of
%its irreducible $(1/2,2)\oplus (2,1/2)$ subspace.
% In pinning down that very subspace by static Lorentz-ivariant projectors,
%the separation of spin-$3/2$ and $5/2$ in $(1/2,2)\oplus (2,1/2)$ can then be
%carried out by means of a dynamical Poincar\'e covariant
%projector, thus ending up with a second order Lagrangian.

\end{abstract}


\maketitle
\tableofcontents
%%%%%%%%%%%%%%%%%%%%%%%%%%%%%%%%%%%%%%%%%%%%%%%%%%%%%%%%%%%%%%%%%%%%%%%%%%%%%%%%%%%%%%%%%%%%%%%%%%%%%%%%%%%%%%%%%%%%%%%
%%%%%%%%%%%%%%%%%%%%%%%%%%%%%%%%%%%%%%%%%%%%%%%%%%%%%%%%%%%%%%%%%%%%%%%%%%%%%%%%%%%%%%%%%%%%%%%%%%%%%%%%%%%%%%%%%%%%%%%
%%%%%%%%%%%%%%%%%%%%%%%%%%%%%%%%%%%%%%%%%%%%%%%%%%%%%%%%%%%%%%%%%%%%%%%%%%%%%%%%%%%%%%%%%%%%%%%%%%%%%%%%%%%%%%%%%%%%%%%
\section{Introduction}\label{sec1}
%%%%%%%%%%%%%%%%%%%%%%%%%%%%%%%%%%%%%%%%%%%%%%%%%%%%%%%%%%%%%%%%%%%%%%%%%%%%%%%%%%%%%%%%%%%%%%%%%%%%%%%%%%%%%%%%%%%%%%%
The theory of fields with spins $s\geq 1$ is mainly based on Lorentz-algebra representation spaces
of multiple spins (and parities)  which are of the type $(j,j)$ for bosons with spin $s=2j$, and of 
the type  $(j,j)\otimes \left[(1/2,0)\oplus (0,1/2)\right]$ for fermions with spin $s=(2j+1/2)$ 
\cite{Weinberg:1995mt}. The particles of interest are associated with the 
highest spins in the spaces under discussion, while their lower spin companions have to be projected 
out in order to ensure the correct number of physical degrees of freedom for spin-$s$ description.
Frequently the lower spin sectors in the aforementioned representation spaces are termed to as 
unphysical, perhaps more in the sense of their exclusion from the physics of the high-spins of prime 
interest than literally in the  sense of non-observability. On the other side, in  recent years strategies have been developed of
keeping the redundant components
in the course of evaluation of a process and eventually removing them only
after the calculation of the observables. Specifically,
relevance of the lower-spins within this context has been noticed
in  the dressing of the spin-$3/2$ propagator within the framework of the
Schwinger-Dyson equation
\cite{Kaloshin:2011gr}, \cite{Kaloshin:2003MPLA}.
Moreover, in Compton scattering off the $\Delta(1232)$ resonance
\cite{Sholten},
they  seem to provide contributions to the non-resonant background
and give rise to interference effects. Finally, in theories 
of supergravity one encounters a metric and a four-vector-spinor Rarita-Schwinger field  that can 
contain fields of lower spins as well \cite{Freed}. For all these reasons clarifying the status of the 
lower spin sector in the spin-$3/2$ Rarita-Schwinger field appears timely. It is the goal of the 
present work to contribute to that clarification. 

The spin-$3/2$ Rarita-Schwinger field is Lorentz transformed as a four-vector with Dirac-spinor 
components, $\psi_\mu\sim (1/2,1/2)\otimes \left[\left(1/2,0 \right)\oplus \left(0,1/2\)\]$.
This Lorentz invariant representation space is  reducible according to 
\aeq
~(1/2,1/2)\otimes [(1/2,0)\oplus (0,1/2)]\longrightarrow
\left[(1/2,0)\oplus (0,1/2)\right]\oplus \left[(1/2,1)\oplus (1,1/2) \right],
\label{RS_red}
\ceq
and the challenge is to construct the space under discussion in such a way that it guarantees that 
the first and second spin-$1/2$ companions to spin-$3/2$ correctly transform as $(1/2,0)\oplus (0,1/2)$, 
and $(1/2,1)\oplus (1,1/2)$, respectively. To achieve our goal we exploit the properties of the 
Casimir invariants of the Lorentz algebra to distinguish between the above irreducible representation 
spaces \cite{Wyborne} and construct corresponding  projectors which bring the great advantage of being
momentum-independent, and which efficiently separate the pure-spin-$1/2$ sector from $(1/2,1)\oplus (1,1/2)$. 
Then the spin-$1/2$ residing in the latter representation space is separated from its spin-$3/2$ companion 
by means of the second order Poincar\'e covariant projector from \cite{Napsuciale:2006wr}, an operator 
based on the two Casimir invariants of the Poincar\'e algebra, the squared momentum, and the squared 
Pauli-Lubanski vector. In so doing, we find two spin-$1/2$ wave equations which we show to describe causal 
propagation of the respective wave fronts in the presence of an electromagnetic field. We furthermore 
calculate Compton scattering off both states, and in finding that the amplitudes corresponding to the first 
spin-$1/2$ are identical to those of a genuine Dirac particle,  conclude on the observability of this very state.
Also for the second one we find finite cross sections in all directions in the ultrarelativistic limit,
and conclude that its observability is not excluded neither by causality of propagation within an 
electromagnetic environment, nor by  unitarity of the Compton scattering  amplitudes.\\

\noindent
The paper is organized as follows. In the next section we present the construction of the momentum-independent
Lorentz projectors and briefly review the recently developed Poincar\'e covariant projector method that has proved
successful in the consistent description of various processes \cite{DelgadoAcosta:2013}--\cite{Napsuciale:2007ry} 
including particles with spins ranging from $1/2$ to $3/2$. There we furthermore present the combined Lorentz- and 
Poincar\'e invariant projectors, develop the formalism for spin-1/2 description in $\psi_\mu$ and obtain the 
corresponding wave equations. In section \ref{sec3} we solve the spin-$1/2$ wave equations from the previous section. 
In section \ref{sec4} we test the causality of the emerging wave equations describing  the spin-$1/2$ degrees of freedom in 
$\psi_\mu$, obtain the associated Lagrangians and describe the electromagnetic properties of the lower spins in 
the four-vector spinor. Section \ref{sec5} is devoted to the calculation of Compton scattering off spin-1/2 transforming as 
$(1/2,0)\oplus (0,1/2)$, or $(1/2,1)\oplus (1,1/2)$. The paper closes with brief conclusions.

%%%%%%%%%%%%%%%%%%%%%%%%%%%%%%%%%%%%%%%%%%%%%%%%%%%%%%%%%%%%%%%%%%%%%%%%%%%%%%%%%%%%%%%%%%%%%%%%%%%%%%%%%%%%%%%%%%%%%%%
%%%%%%%%%%%%%%%%%%%%%%%%%%%%%%%%%%%%%%%%%%%%%%%%%%%%%%%%%%%%%%%%%%%%%%%%%%%%%%%%%%%%%%%%%%%%%%%%%%%%%%%%%%%%%%%%%%%%%%%
%%%%%%%%%%%%%%%%%%%%%%%%%%%%%%%%%%%%%%%%%%%%%%%%%%%%%%%%%%%%%%%%%%%%%%%%%%%%%%%%%%%%%%%%%%%%%%%%%%%%%%%%%%%%%%%%%%%%%%%
\section{The method of the combined Lorentz- and Poincar\'e invariant projectors}\label{sec2}
%%%%%%%%%%%%%%%%%%%%%%%%%%%%%%%%%%%%%%%%%%%%%%%%%%%%%%%%%%%%%%%%%%%%%%%%%%%%%%%%%%%%%%%%%%%%%%%%%%%%%%%%%%%%%%%%%%%%%%%
Equations of motion for particles transforming according to a given Lorentz-algebra representation space,
$\lbrace \kappa \rbrace$ spanned by the generic degrees of freedom $\psi({\mathbf p},\lambda)$, with  ${\mathbf p}$ 
denoting the three-dimensional momentum, and $\lambda$ standing for the set of quantum numbers characterizing the degrees 
of freedom under consideration, are straightforwardly constructed from a projector (see \cite{DelgadoAcosta:2013} for 
more details), call it $\Pi_{\lbrace \kappa\rbrace }({\mathbf p})$, as 
\begin{equation}
\Pi_{\lbrace \kappa \rbrace }({\mathbf p})\psi({\mathbf
p},\lambda)=\psi({\mathbf p},\lambda).
\label{prj_eqs}
\end{equation}
In single-spin valued representation spaces, such as the Dirac one, it is customary to use for 
$\Pi_{\lbrace \kappa \rbrace }({\mathbf p},\lambda )$ one of the  parity projectors, $(\pm\not{p}+m)/2m$. On the contrary, 
in order to track down the desired  spin-$s_i$ within an irreducible  representation space containing $N$ spin-sectors, 
$s_1,...,s_N$, the  projector $\Pi^{s_i}_{\lbrace \kappa\rbrace }({\mathbf p},\lambda )$ has to be taken as  the  product 
of $(N-1)$  covariant spin-projectors of the type  designed in refs.~\cite{DelgadoAcosta:2013}--\cite{Napsuciale:2007ry}
on the basis of the two Casimir invariants of the Poincar\'e group, the squared four-momentum, $P^2$, and the squared 
Pauli-Lubanski vector, $W^2$. The disadvantage is that in so doing one ends up with uncomfortable to deal with
equations of the order $p^{2(N-1)}$. In the present study we reveal the advantages of describing spin-$s$ in terms of
sufficiently large reducible representation spaces equipped by separate Lorentz- and Dirac spinor indices and such that
the desired spin is contained either within  a single-, or within a maximally two-spin valued irreducible subspace.
The two spins within the latter subspace can be separated by a Poincar\'e covariant projector, which is second order in 
the momenta. All the remaining subspaces can be distinguished by the Casimir invariants of the Lorentz  algebra and removed 
by properly constructed momentum-independent projectors. In effect, $\Pi_{\lbrace \kappa\rbrace}({\mathbf p},\lambda )$
can be furnished as a product of a single Poincar\'e covariant projector that is of second order in the momenta  and 
several momentum independent Lorentz projectors which results in quadratic wave equations for any spin. Below we shall 
illustrate the above concept on the example of  the four-vector spinor Rarita-Schwinger representation space that is of 
prime interest to the present study. Our point is that the Lorentz-invariant projectors are indispensable for the correct 
identification of the spin-$1/2$ degrees of freedom in $\psi_\mu$ that transform irreducibly under Lorentz transformations.
%%%%%%%%%%%%%%%%%%%%%%%%%%%%%%%%%%%%%%%%%%%%%%%%%%%%%%%%%%%%%%%%%%%%%%%%%%%%%%%%%%%%%%%%%%%%%%%%%%%%%%%%%%%%%%%%%%%%%%%
%%%%%%%%%%%%%%%%%%%%%%%%%%%%%%%%%%%%%%%%%%%%%%%%%%%%%%%%%%%%%%%%%%%%%%%%%%%%%%%%%%%%%%%%%%%%%%%%%%%%%%%%%%%%%%%%%%%%%%%
\subsection{Poincar\'e covariant spin-$s$ projectors }\label{sec2a}
%%%%%%%%%%%%%%%%%%%%%%%%%%%%%%%%%%%%%%%%%%%%%%%%%%%%%%%%%%%%%%%%%%%%%%%%%%%%%%%%%%%%%%%%%%%%%%%%%%%%%%%%%%%%%%%%%%%%%%%
The Poincar\'e covariant projector method \cite{Napsuciale:2006wr} relies upon
the two Casimir operators of the Poincar\'e algebra, the
squared four-momentum, $P^2$, and the squared Pauli-Lubanski vector,  $W^2$.
These operators fix in their turn the
 mass-$m$,  and the  spin-$s$ quantum numbers of the states, here denoted by
$w^{(m,s)}$, transforming according to the  representation space of interest.
One has
\aec
P^2 w^{(m,s)}&=&m^2 w^{(m,s)},\\
W^2 w^{(m,s)}&=&-p^2 s(s+1)w^{(m,s)}.
\label{P2_W2}
\cec
The $W^2$ Casimir invariant is constructed from the elements,
$M_{\m\n}$, of the Lorentz algebra in the representation of interest.
Specifically for the four-vector spinor one finds,
\aec
~[W^2]_{\a\b}&=&[W^\m]_{\a}{}^\g [W_\m]_{\g\b}= [T_{\m\n}]_{\a\b} p^\m
p^\n\label{w2rs},\\
~[T_{\m\n}]_{\a\b}&=&\frac{1}{4}\e^{\s}{}_{\l\t\m}\e_{\s\h\z\n}[M^{\l\t}]_{\a}{}^\d
[M^{\h\z}]_{\d\b},\label{tw2rs}\\
~[M_{\m\n}]_{\a\b}&=&[M^V_{\m\n}]_{\a\b}+g_{\a\b} M^S_{\m\n}\label{gensrs}.
\cec
where $[M^V_{\m\n}]_{\a\b}$ and $[M^S_{\m\n}]$ represent  the Lorentz-group  generators within
the respective four-vector--,  and the Dirac-spinor building blocks. Their explicit forms read
\aec
~[M^V_{\m\n}]_{\a\b}&=&i(g_{\a\m}g_{\b\n}-g_{\a\n}g_{\b\m}),\label{gensv}\\
M^S_{\m\n}&=&\frac{1}{2}\s_{\m\n}=\frac{i}{4}[\g_\m,\g_{\n}]\label{genss},
\cec
with $\g_\m$ being the standard  Dirac  matrices. Notice that the operators in
\eqref{w2rs},\eqref{tw2rs} and
\eqref{gensrs} are $16\times 16$ matrices just as are the generators in
$\psi_\mu$, and consequently carry  next to the Lorentz indices, also spinor
indices, here suppressed  for the sake of simplifying notations.
The Rarita-Schwinger  representation contains only
two spin sectors, namely $s_1=1/2$ and $s_2=3/2$, and one can construct projectors,
$\mathcal{P}^{(m,s)}_{W^2}$, over spin and
mass in terms of the $P^2$ and $W^2$ operators as follows (see
\cite{Napsuciale:2006wr} for the details):
\aec
\mathcal{P}^{(m,1/2)}_{W^2}w^{(m,1/2)}&=&\frac{P^2}{m^2}\(\frac{W^2-\e_{3/2}}{\e_{1/2}-\e_{3/2}}\)w^{(m,1/2)}
=w^{(m,1/2)},\label{w2eq12}\\
\mathcal{P}^{(m,3/2)}_{W^2}w^{(m,3/2)}&=&\frac{P^2}{m^2}\(\frac{W^2-\e_{1/2}}{\e_{3/2}-\e_{1/2}}\)w^{(m,3/2)}
=w^{(m,3/2)}.\label{w2eq32}
\cec
Here, $\e_s=-p^2 s(s+1)$ is the $W^2$ eigenvalue corresponding to $w^{(m,s)}$,
the mass-$m$ and spin-$s$ eigenstates to the operators $P^2$ and $W^2$.
With the aid of
\eqref{w2rs}--\eqref{genss}
we find
\aec
~[\mathcal{P}^{(m,1/2)}_{W^2}(p)]_{\a\b}&=&\frac{1}{3m^2}\([T_{\m\n}]_{\a\b}+\frac{15}{4}
g_{\a\b}g_{\m\n}\)p^\m p^\n
=\frac{p^2}{m^2}[\mathbb{P}^{(1/2)}(p)]_{\a\b},
\label{pro_12}\\
~[\mathcal{P}^{(m,3/2)}_{W^2}(p)]_{\a\b}&=&-\frac{1}{3m^2}\([T_{\m\n}]_{\a\b}+\frac{3}{4}
g_{\a\b}g_{\m\n}\)p^\m p^\n
=\frac{p^2}{m^2}[\mathbb{P}^{(3/2)}(p)]_{\a\b},
\label{pro_32}
\cec
where, as it  will be made explicit  later, the operators
$[\mathbb{P}^{(s)}(p)]_{\a\b}$ project over states of spin-$s$.
Expressions for the latter  can be found, among others, in Ref.~\cite{VanNieuwenhuizen:1981ae} and read:
\aec
\left[\mathbb{P}^{(1/2)}(p) \right]_{\a\b}
&=&\frac{1}{3}\g_\a \g_\b+\frac{1}{3p^2}(\not{p}\g_\a p_\b+p_\a \g_\b
\not{p})\label{p12},\\
\left[\mathbb{P}^{(3/2)}(p) \right]_{\a\b}
&=&g_{\a\b}-\frac{1}{3}\g_\a \g_\b-\frac{1}{3p^2}(\not{p}\g_\a p_\b+p_\a \g_\b
\not{p})\label{p32}.
\cec
The equation of motion for spin-$s$ in $\psi_\mu$  resulting
from \eqref{prj_eqs},  \eqref{w2eq12}, and \eqref{w2eq32}  is
\aec
~ \left([\mathcal{T}^{(s)}_{\m\n}]_{\a\b}p^\m p^\n-m^2
g_{\a\b}\right)[w^{(s)}]^\b&=&0, \quad s=\frac{1}{2},\frac{3}{2},\label{teqrs}
\cec
where
\aec
~[\mathcal{T}^{(1/2)}_{\m\n}]_{\a\b}&=&\frac{1}{3}\([T_{\m\n}]_{\a\b}+\frac{15}{4}
g_{\a\b}g_{\m\n}\),\label{tp12}\\
~[\mathcal{T}^{(3/2)}_{\m\n}]_{\a\b}&=&-\frac{1}{3}\([T_{\m\n}]_{\a\b}+\frac{3}{4}
g_{\a\b}g_{\m\n}\)\label{tp13}.
\cec
Carrying out the  contractions amounts to  the following free equations of motion:
\aec
\(\frac{1}{3}\g_\a \g_\b p^2 +\frac{1}{3}(\not{p}\g_\a p_\b+p_\a\g_\b\not{p})-m^2
g_{\a\b}\)[w^{(1/2)}]^\b
&=&0,\label{s12feq}\\
\(-\frac{1}{3}\g_\a \g_\b p^2-\frac{1}{3}(\not{p}\g_\a p_\b+p_\a
\g_\b\not{p})-(m^2-p^2)g_{\a\b}\)[w^{(3/2)}]^\b
&=&0.\label{s32feq}
\cec
Notice that the Poincar\'e  covariant
projector method designs  the spin-$s$ description on the basis of
the transformation properties of the free particles.
It therefore leaves their electromagnetic properties unspecified
and allows to replace  $[\mathcal{T}^{(s)}_{\m\n}]_{\a\b}$ by tensors which
are  equivalent on-shell, though become
distinct upon introducing interactions.
The electromagnetic constants   have  to be fixed
at a later stage by some properly chosen dynamical constraints.
As already announced  in the previous section, from now on we focus
on the two spin-$1/2$ sectors in $\psi_\mu$.
%%%%%%%%%%%%%%%%%%%%%%%%%%%%%%%%%%%%%%%%%%%%%%%%%%%%%%%%%%%%%%%%%%%%%%%%%%%%%%%%%%%%%%%%%%%%%%%%%%%%%%%%%%%%%%%%%%%%%%%
%%%%%%%%%%%%%%%%%%%%%%%%%%%%%%%%%%%%%%%%%%%%%%%%%%%%%%%%%%%%%%%%%%%%%%%%%%%%%%%%%%%%%%%%%%%%%%%%%%%%%%%%%%%%%%%%%%%%%%%
\subsection{Momentum independent Lorentz invariant projectors on irreducible representation spaces}\label{sec2b}
%%%%%%%%%%%%%%%%%%%%%%%%%%%%%%%%%%%%%%%%%%%%%%%%%%%%%%%%%%%%%%%%%%%%%%%%%%%%%%%%%%%%%%%%%%%%%%%%%%%%%%%%%%%%%%%%%%%%%%%
The spin-$1/2$ projectors composed in the preceding section from the Casimir
invariants of the Poincar\'e algebra are indifferent
to the reducibility of the Lorentz representation and one can not expect that
their eigenvectors transform
irreducibly. To ensure the correct Lorentz transformation properties of the
two spin-$1/2$ degrees of freedom in $\psi_\mu$
it is necessary to construct additional projectors based upon  the Casimir invariants
of the Lorentz algebra itself.
The Lorentz algebra has two Casimir operators, usually  denoted by $F$
and $G$, and given by \cite{Wyborne}
\aec
~[F]_{\a\b}&=&\frac{1}{4}[M^{\m\n}]_\a{}^\g
[M_{\m\n}]_{\g\b}=\frac{9}{4}g_{\a\b}+\frac{i}{2}\s_{\a\b},\\
~[G]_{\a\b}&=&\frac{1}{4}\e_{\m\n\r\s}[M^{\m\n}]_\a{}^\g
[M^{\r\s}]_{\g\b}=\g^5 \s_{\a\b}-\frac{3}{2}i\g^5 g_{\a\b},
\cec
where use has been made of  \eqref{gensrs} to simplify the expressions.
Their respective eigenvalue problems read,
\aec
F \,w_{(j_1,j_2)}&=&\frac{1}{2}(K(K+2)+M^2)w_{(j_1,j_2)},\nonumber\\
G \,w_{(j_1,j_2)}&=&i M(K+1)w_{(j_1,j_2)},
\label{FCasm}
\cec
where $w_{(j_1,j_2)}$ are the states transforming irreducibly as
$(j_2,j_1)\oplus(j_1,j_2)$, and
\aeq
K=j_1+j_2,\qquad M=\vert j_1-j_2 \vert.
\ceq
The $F$ eigenvectors are of well defined parities, while those of
$G$ are chiral states. In the following we choose to work with the $F$ invariant.
We have verified that the $F$ invariant commutes with
the $W^2$-operator of the squared Pauli-Lubanski vector, thus providing
the possibility to design a Lorentz  projector whose eigenvectors
are simultaneously  eigenvectors to the Poincar\'e covariant spin-$1/2$ projector
in \eqref{p12}.
Such projectors, here denoted by $\mathcal{P}_F^{(j_1)}$,
with $j_1=0,1$, select  $w^{(j_1)}$ states transforming according to
the $(1/2,j_1)\oplus(j_1,1/2)$ sectors of the Rarita-Schwinger  representation space:
\aec
\mathcal{P}^{(0)}_F
w^{(0)}&=&\(\frac{F-\l_{1}}{\l_{0}-\l_{1}}\)w^{(0)}=w^{(0)},\label{feq12}\\
\mathcal{P}^{(1)}_F
w^{(1)}&=&\(\frac{F-\l_{0}}{\l_{1}-\l_{0}}\)w^{(1)}=w^{(1)}.\label{feq32}
\cec
Here $\l_{j_1}$ are the $F$ eigenvalues 
\aeq\label{evF}
\l_0=\frac{3}{4},\qquad 
\l_1=\frac{11}{4},
\ceq
corresponding to $j_1=0,1$ and $j_2=1/2$. The projector operators  are then easily calculated as,
\aec
~[\mathcal{P}^{(0)}_F]_{\a\b}&=&\frac{1}{4}\g_\a \g_\b,\label{pf0}\\
~[\mathcal{P}^{(1)}_F]_{\a\b}&=&g_{\a\b} -\frac{1}{4}\g_\a \g_\b\label{pf1}.
\cec
The orthogonality and completeness properties of the above set of operators
are easily seen.
Compared to the Poincar\'e covariant projectors, the Lorentz ones have the
advantage to be  momentum-independent,
which will allow us to remove redundant irreducible sectors of a reducible
representation space without
increasing the power of the momentum dependence  of the wave equation.
Finally, it is not difficult to verify that
\aec
~[\mathcal{P}^{(0)}_F]_{\a}{}^\g [\mathcal{P}^{(m,3/2)}_{W^2}(p)]_{\g\b}&=&0,\\
~[\mathcal{P}^{(1)}_F]_{\a}{}^\g [\mathcal{P}^{(m,3/2)}_{W^2}(p)]_{\g\b}&=&[\mathcal{P}^{(m,3/2)}_{W^2}(p)]_{\a\b},
\cec
which confirms the correct assignment of  spin-$3/2$ to the
 $(1/2,1)\oplus(1,1/2)$ invariant subspace in the four-vector--spinor.

The next section is devoted to the solution of the above Eqs.~\eqref{feq12}
and \eqref{feq32}.
%%%%%%%%%%%%%%%%%%%%%%%%%%%%%%%%%%%%%%%%%%%%%%%%%%%%%%%%%%%%%%%%%%%%%%%%%%%%%%%%%%%%%%%%%%%%%%%%%%%%%%%%%%%%%%%%%%%%%%%
%%%%%%%%%%%%%%%%%%%%%%%%%%%%%%%%%%%%%%%%%%%%%%%%%%%%%%%%%%%%%%%%%%%%%%%%%%%%%%%%%%%%%%%%%%%%%%%%%%%%%%%%%%%%%%%%%%%%%%%
\subsection{Combined Lorentz-- and Poincar\'e invariant  projectors }\label{sec2c}
%%%%%%%%%%%%%%%%%%%%%%%%%%%%%%%%%%%%%%%%%%%%%%%%%%%%%%%%%%%%%%%%%%%%%%%%%%%%%%%%%%%%%%%%%%%%%%%%%%%%%%%%%%%%%%%%%%%%%%%

The idea of the present work is to identify the lower spin degrees of freedom
in $\psi_\mu$ by means of the following
combined Lorentz- and Poincar\'e invariant projectors:
\aec
\Pi _{(1/2,j_1)\oplus(j_1,1/2)}w^{(m,s)}_{(j_1)}&=&w^{(m,1/2)}_{(j_1)},
\quad \Pi _{(1/2,j_1)\oplus(j_1,1/2)}= \mathcal{P}^{(j_1)}_F\,\mathcal{P}^{(m,1/2)}_{W^2},
\label{comeq}
\cec
which translates into the following two equations of motion (compressed in one):
\aeq\label{eomsj1}
\([\G^{(j_1)}_{\m\n}]_{\a\b}p^\m p^\n-m^2 g_{\a\b}\)[w^{(m,1/2)}_{(j_1)}]^\b=0,
\quad j_1=0,1.
\ceq
Here, the Lorentz-- and Poincar\'e- projectors enter the
$[\G^{(j_1)}_{\m\n}]_{\a\b}$ tensors as
\aeq
~[\G^{(j_1)}_{\m\n}]_{\a\b}p^\m p^\n= m^2
[\mathcal{P}^{(j_1)}_F]_{\a}{}^\g [\mathcal{P}^{(m,1/2)}_{W^2}]_{\g\b}
=[\mathcal{P}^{(j_1)}_F]_{\a}{}^\g[\mathcal{T}^{(1/2)}_{\m\n}]_{\g\b}p^\m
p^\n,
\ceq
with $[\mathcal{T}^{(1/2)}_{\m\n}]_{\g\b}$ taken from \eqref{tp12}.
Carrying out the contraction in the $\mu$ and $\nu$ indices,
the equation for  $(1/2,0)\oplus(0,1/2)$ emerges
particularly simple as,
\aeq
\(\frac{1}{4}\g_\a \g_\b p^2-m^2 g_{\a\b}\)[w^{(m,1/2)}_{(0)}]^\b=0,
\label{dir_12}
\ceq
while the one for   $(1/2,1)\oplus(1,1/2)$ results as,
\aeq
\[\frac{4}{3}\(p_\a-\frac{1}{4}\g_\a\not{p}\)\(p_\b-\frac{1}{4}\not{p}\g_\b\)-m^2
g_{\a\b}\][w^{(m,1/2)}_{(1)}]^\b=0.\label{ceq1}
\ceq
The latter equation incorporates  an auxiliary condition as visible by contracting
it  by $\g^\a$, finding
\aeq
-m^2\g_\b [w^{(m,1/2)}_{(1)}]^\b=0.\label{auxj11}
\ceq
The meaning of \eqref{auxj11} is that  $w^{(m,1/2)}_{(1)}$ does not have any
projection on
$(1/2,0)\oplus(0,1/2)$, as it should be, and in accord with the  established
reduction of
the Rarita-Schwinger space, discussed in the introduction.
We furthermore  define the following pair of orthogonal matrices,
\aec
~[f^{(0)}(p)]^\a&=&\frac{1}{2m}\g^\a \not{p},\label{ef0}\\
~[f^{(1)}(p)]^\a&=&\frac{2}{\sqrt{3}m}\(p^\a-\frac{1}{4}\g^\a\not{p}\)\label{ef1},
\cec
which are orthonormalized on mass-shell according to,
\aeq\label{eqforty}
~[\overline{f}^{(j)}(p)]^\a[f^{(j')}(p)]_\a=\d^j_{j'}\frac{p^2}{m^2},
\ceq
with
$[\overline{f}^{(j_1)}(\mathbf{p})]^\a=\g^0([f^{(j_1)}(\mathbf{p})]^\a)^\dagger\g^0$.
In terms of the
$[f^{(j_1)}(p)]^\a$ matrices, the kinetic terms of the equations of motion are bi-linearized according to,
\aeq\label{fequations}
\([f^{(j_1)}(p)]_\a [\overline{f}^{(j_1)}(p)]_\b -
g_{\a\b}\)[w^{(m,1/2)}_{(j_1)}]^\b=0.
\ceq
In combination with the Eq. \eqref{eqforty} one sees that the $[f^{(j_1)}(p)]^\a$ matrices take the part in \eqref{dir_12} and \eqref{ceq1}, i.e. in the spin-1/2 sector of $\y_\m$, of the Feynman slash, $p\!\!\!/$, in the bi-linearization of the Klein-Gordon equation where $p\!\!\!/\cdot p\!\!\!/=p^2$. Now the propagator for each $j_1$-value is then the inverse of the respective equation operator,
\aeq\label{props0}
~[S^{(j_1)}(p)]_{\a\b}=\([\G^{(j_1)}_{\m\n}]_{\a\b}p^\m p^\n-m^2 g_{\a\b}\)^{-1},
\ceq
and given by
\aeq\label{props}
~[S^{(j_1)}(p)]_{\a\b}=\frac{[\Delta^{(j_1)}(p)]_{\a\b}}{p^2-m^2+i\e},
\ceq
where
\aec
~[\D^{(0)}(p)]_{\a\b}&=&\frac{1}{m^2}\(\frac{1}{4}p^2\g_\a \g_\b-(p^2-m^2)g_{\a\b}\),\label{deltaj0}\\
~[\D^{(1)}(p)]_{\a\b}&=&\frac{1}{m^2}\[\frac{4}{3}\(p_\a-\frac{1}{4}\g_\a\not{p}\)\(p_\b-\frac{1}{4}\not{p}\g_\b\)+(m^2-p^2)g_{\a\b}\]\label{deltaj1}.
\cec
In terms of the  Lorentz- and Poincar\'e- projectors this is equivalent to,
\aec
~[\D^{(j_1)}(p)]_{\a\b}&=&\frac{p^2}{m^2}[\mathcal{P}^{(j_1)}_F]_{\a}{}^\g[\mathbb{P}^{(1/2)}(p)]_{\g\b}+\frac{(p^2-m^2)}{m^2}g_{\a\b},
\cec
while in terms of the $[f^{(j_1)}(p)]^\a$ matrices this is just
\aec
~[\D^{(j_1)}(p)]_{\a\b}&=&[f^{(j_1)}(p)]_\a
[\overline{f}^{(j_1)}(p)]_\b+\frac{(p^2-m^2)}{m^2}g_{\a\b}.
\cec
%%%%%%%%%%%%%%%%%%%%%%%%%%%%%%%%%%%%%%%%%%%%%%%%%%%%%%%%%%%%%%%%%%%%%%%%%%%%%%%%%%%%%%%%%%%%%%%%%%%%%%%%%%%%%%%%%%%%%%%
%%%%%%%%%%%%%%%%%%%%%%%%%%%%%%%%%%%%%%%%%%%%%%%%%%%%%%%%%%%%%%%%%%%%%%%%%%%%%%%%%%%%%%%%%%%%%%%%%%%%%%%%%%%%%%%%%%%%%%%
%%%%%%%%%%%%%%%%%%%%%%%%%%%%%%%%%%%%%%%%%%%%%%%%%%%%%%%%%%%%%%%%%%%%%%%%%%%%%%%%%%%%%%%%%%%%%%%%%%%%%%%%%%%%%%%%%%%%%%%
\section{Solutions to the spin-1/2 equations of motion}\label{sec3}
%%%%%%%%%%%%%%%%%%%%%%%%%%%%%%%%%%%%%%%%%%%%%%%%%%%%%%%%%%%%%%%%%%%%%%%%%%%%%%%%%%%%%%%%%%%%%%%%%%%%%%%%%%%%%%%%%%%%%%%
In the current section we compare the spin-$1/2$ solutions of the pure Poincar\'e covariant projector in
\eqref{s12feq}  to those of the combined Lorentz- and Poincar\'e projectors in \eqref{dir_12}--\eqref{ceq1} 
and draw some non-trivial conclusions regarding their space-time properties.
%%%%%%%%%%%%%%%%%%%%%%%%%%%%%%%%%%%%%%%%%%%%%%%%%%%%%%%%%%%%%%%%%%%%%%%%%%%%%%%%%%%%%%%%%%%%%%%%%%%%%%%%%%%%%%%%%%%%%%%
%%%%%%%%%%%%%%%%%%%%%%%%%%%%%%%%%%%%%%%%%%%%%%%%%%%%%%%%%%%%%%%%%%%%%%%%%%%%%%%%%%%%%%%%%%%%%%%%%%%%%%%%%%%%%%%%%%%%%%%
\subsection{The eigenvectors to the Poincar\'e covariant spin-$1/2$  projection}\label{sec3a}
%%%%%%%%%%%%%%%%%%%%%%%%%%%%%%%%%%%%%%%%%%%%%%%%%%%%%%%%%%%%%%%%%%%%%%%%%%%%%%%%%%%%%%%%%%%%%%%%%%%%%%%%%%%%%%%%%%%%%%%
The solutions of the eigenvalue problem of the spin-$1/2$ Poincar\'e projector in \eqref{w2eq12}
are no more but
the  spin-$1/2$ eigenvectors of $W^2$, which we here denote by
$[w({\mathbf p},\l)]^\a$.  They emerge in the direct product of the
four vectors in $(1/2,1/2)$ and the spinors in $(1/2,0)\oplus(0,1/2)$.
The $(1/2,1/2)$ representation is spanned by one scalar and three vectorial
degrees of freedom.
The spin-$1/2$ four-vector spinors emerging from the coupling of the scalar in
$(1/2,1/2)$ to the Dirac spinor
will be termed to as scalar-spinors ($SS$),
\aeq
[w^{SS}_{\pm}({\mathbf p},\l)]^\a=\f^\a(p) u_{\pm}({\mathbf p},\l)=\frac{p^\a}{m}
u_{\pm}({\mathbf p},\l).\label{ssdef}
\ceq
Here, $\f^\a(p)=p^\a/m$ is the only spin-$0$ vector in $(1/2,1/2)$, while
$u_{\pm}({\mathbf p},\l)$ are the usual  Dirac spinors in $(1/2,0)\oplus(0,1/2)$,
of positive ($+$) and negative ($-$) parities,
(i.e., $u$-- and $v$-spinors in the terminology of \cite{Bjorken}), and whose polarizations
are $\l=1/2,-1/2$.
The spin-$1/2$ vectors emerging of the coupling of the spin-$1^-$ vectors in
$(1/2,1/2)$, in the cartesian basis denoted by $\h^\a({\mathbf p},\ell)$ 
(with $\ell=-1,0,1$) \cite{Napsuciale:2006wr}, to the Dirac spinor 
will be termed to  as vector-spinors ($VS$). They are constructed within the ordinary angular momentum
coupling scheme in terms of appropriate Clebsch-Gordan coefficients, and read
\aec
~[w^{VS}_{\pm}({\mathbf p},1/2)]^\a&=&-\sqrt{\frac{1}{3}}[\h({\mathbf p},0)]^{\a}u_{\mp}({\mathbf p},1/2)+\sqrt{\frac{2}{3}}[\h({\mathbf p},1)]^\a
u_\mp({\mathbf p},-1/2)\label{p12vs},\\
~[w^{VS}_{\pm}({\mathbf p},-1/2)]^\a&=&\sqrt{\frac{1}{3}}[\h({\mathbf p},0)]^{\a}u_{\mp}({\mathbf p},-1/2)-\sqrt{\frac{2}{3}}[\h({\mathbf p},-1)]^\a
u_\mp({\mathbf p},1/2).\label{m12vs}
\cec
Notice that while $\f^\a(p)$ is of positive parity, the parity of
$[\h({\mathbf p},\ell)]^\a$
is negative. In consequence,
the positive- (negative-) parity scalar-spinors are made up of positive-
(negative-) parity Dirac spinors, while
the positive- (negative-) parity vector-spinors are made up of negative-
(positive-) parity Dirac spinors.
Alternatively, the above vector-spinors can be derived in exploiting the following
relationship,
\aeq
~[W^2(p)]_{\a}{}^\b
[W^S(p)]_\b=-p^2\frac{1}{2}\(\frac{1}{2}+1\)[W^S(p)]_\a,\label{w2wsrel}
\ceq
where $[W^S(p)]_\a$ is the Pauli-Lubanski operator in the Dirac-spinor representation,
\aeq
~[W^S(p)]^\m=\frac{1}{2}\e^{\m\n\s\r}M^S_{\s\r}p_\n=-\frac{i}{2}\g^5
\s^{\m\n}p_\n.
\ceq
The relationship in \eqref{w2wsrel} suggests that one can construct spin-$1/2$
eigenstates to
$[W^2(p)]_{\a\b}$ by the aid of the operator $[W^S(p)]_\a$ of the
Pauli-Lubanski vector.
Indeed, one can verify that the vector-spinors in the equations \eqref{p12vs} and
\eqref{m12vs}   are equivalent to
\aeq
~[w^{VS}_{\pm}({\mathbf p},\l)]^\a=\frac{2}{\sqrt{3}m}[W^S(p)]_\a \g^5
u_{\pm}({\mathbf p},\l).\label{vsdef}
\ceq
The definitions in \eqref{ssdef} and \eqref{vsdef} will proof very useful in the
following. Specifically, the
demonstration of the orthogonality between scalar- and vector-spinors will benefit from the
well known property of the Pauli-Lubanski vector of being divergence-less,
\aeq
p^\a [W^S(p)]_\a=0,
\ceq
which  implies  the following  condition satisfied by the vector-spinors:
\aeq
p^\a [w^{VS}_{\pm}({\mathbf p},\l)]_\a=0.
\ceq
In effect, one first finds the four independent scalar-spinors, and then as
another set, the four independent
vector-spinors, summing up  to eight spin-$1/2$ states, as it should be.
Together with  the eight
independent spin-$3/2$ states (not considered here) the
total of sixteen independent degrees of freedom in the four-vector-spinor is
recovered.
%%%%%%%%%%%%%%%%%%%%%%%%%%%%%%%%%%%%%%%%%%%%%%%%%%%%%%%%%%%%%%%%%%%%%%%%%%%%%%%%%%%%%%%%%%%%%%%%%%%%%%%%%%%%%%%%%%%%%%%
\subsubsection{Parity of the states diagonalizing the Poincar\'e covariant spin-$1/2$  projectors }\label{sec3a1}
%%%%%%%%%%%%%%%%%%%%%%%%%%%%%%%%%%%%%%%%%%%%%%%%%%%%%%%%%%%%%%%%%%%%%%%%%%%%%%%%%%%%%%%%%%%%%%%%%%%%%%%%%%%%%%%%%%%%%%%
As long as the spin-$1/2$ states in \eqref{ssdef} and \eqref{vsdef}  have
well defined parities,
one can construct from them the corresponding parity projectors. Each  spin-$1/2$ is normalized according to
\aeq
~[\overline{w}^{SS/SV}_{\pm}(\mathbf{p},\l)]^\a
[w^{SS/SV}_{\pm}(\mathbf{p},\l)]_\a=\pm
1\label{s12snor},
\ceq
where use has been made of
\aec
~[\overline{w}^{SS}_{\pm}(\mathbf{p},\l)]^\a&=&p^\a\overline{u}_{\pm}(\mathbf{p},\l),\label{ssbar}\\
~[\overline{w}^{VS}_{\pm}(\mathbf{p},\l)]^\a&=&\overline{u}_{\pm}(\mathbf{p},\l)\g^5[W^S(\mathbf{p})]_\a\label{vsbar},
\cec
with $\overline{u}_{\pm}(\mathbf{p},\l)=[\g^0
{u}_{\pm}(\mathbf{p},\l)]^\dagger$ as customary. The parity projectors are
then obtained from \eqref{s12snor}-\eqref{vsbar} according to
\aeq\label{ppdef}
~[\mathbb{P}^{SS/SV}_{\pm}(\mathbf{p})]_{\a\b}=\pm\sum_\l
[w^{SS/SV}_{\pm}(\mathbf{p},\l)]_\a
\overline{w}^{SS/SV}_{\pm}(\mathbf{p},\l)]_\b,
\ceq
which amounts to the following explicit  expressions,
\aec\label{proyss}
\[\mathbb{P}_{\pm}^{SS}(\mathbf{p})\]_{\a\b}&=&\frac{\pm
\not{p}+m}{2m}\frac{1}{m^2}p_\a p_\b,\\
\[\mathbb{P}_{\pm}^{VS}(\mathbf{p})\]_{\a\b}&=&\frac{\mp
\not{p}+m}{2m}\frac{1}{3 m^2}\s_{\a\m}\s_{\b\n}p^\m p^\n\label{proyvs}.
\cec
where we have used
\aeq\label{diracproyector}
\sum_{\l}u(\mathbf{p},\l)\overline{u}(\mathbf{p},\l)=\frac{\not{p}+m}{2m}.
\ceq
The projectors in \eqref{proyss} and \eqref{proyvs} can also be cast in terms of the more 
familiar spin-$1/2$ projectors, $[\mathbb{P}^{(1/2)}(p)]_{\a\b}$ from \eqref{p12} and 
\cite{VanNieuwenhuizen:1981ae}:
\aec
~[\mathbb{P}_{11}^{(1/2)}(p)]_{\a\b}&=&-\frac{p_\a p_\b}{p^2}+\frac{1}{3}\g_\a
\g_\b +\frac{1}{p^2}(\not{p}\g_\a p_\b+p_\a\g_\b\not{p}),\\
~[\mathbb{P}_{22}^{(1/2)}(p)]_{\a\b}&=&[\mathbb{P}^{(1/2)}(p)]_{\a\b}-[\mathbb{P}_{11}^{(1/2)}(p)]_{\a\b}=\frac{p_\a
p_\b}{p^2},
\cec
in so doing, we establish the following relationships between parity- and spin
projector operators:
\aec
~\[\mathbb{P}_{\pm}^{SS}(\mathbf{p})\]_{\a\b}&=&\frac{\pm
\not{p}+m}{2m}[\mathbb{P}_{22}^{(1/2)}(\mathbf{p})]_{\a\b},\label{ssp}\\
~\[\mathbb{P}_{\pm}^{VS}(\mathbf{p})\]_{\a\b}&=&\frac{\mp
\not{p}+m}{2m}[\mathbb{P}_{11}^{(1/2)}(\mathbf{p})]_{\a\b}.\label{vsp}
\cec
Notice that we use the boldface  $(\mathbf{p})$ when the mass-shell condition
$p^2=m^2$ holds valid. We can also construct
the so called switch projectors by making combinations between scalar-spinors
and vector-spinors. However, we prefer to sum up  all four projectors in
\eqref{ssp} and \eqref{vsp} with the aim to recover
$[\mathbb{P}^{(1/2)}(p)]_{\a\b}$ in \eqref{pro_12} as previously obtained from the
Poincar\'e covariant spin-$1/2$ projector as
\aeq
\frac{1}{m^2}\(\frac{1}{3}\s_{\a\m}\s_{\b\n}+g_{\a\m}g_{\b\n}\)p^\m
p^\n=[\mathbb{P}^{(1/2)}(\mathbf{p})]_{\a\b}\label{stp},
\ceq
where we have made use of \eqref{ssdef} and \eqref{vsdef}.
The procedure of constructing $[\mathbb{P}^{(1/2)}(p)]_{\a\b}$ from the
parity eigenstates happens to prescribe the correct
$p_\mu p_\nu$ ordering in the momentum  dependence of the equation of motion,
a circumstance that will prove crucial upon gauging.
As a next step we shall obtain  the explicit expressions for the states that
diagonalize the combined  Lorentz- and Poincar\'e invariant projectors.
%%%%%%%%%%%%%%%%%%%%%%%%%%%%%%%%%%%%%%%%%%%%%%%%%%%%%%%%%%%%%%%%%%%%%%%%%%%%%%%%%%%%%%%%%%%%%%%%%%%
%%%%%%%%%%%%%%%%%%%%%%%%%%%%%%%%%%%%%%%%%%%%%%%%%%%%%%%%%%%%%%%%%%%%%%%%%%%%%%%%%%%%%%%%%%%%%%%%%%%
\subsection{The spin-$1/2$ eigenstates to the combined Lorentz- and Poincar\'e
invariant projectors}\label{sec3b}
%%%%%%%%%%%%%%%%%%%%%%%%%%%%%%%%%%%%%%%%%%%%%%%%%%%%%%%%%%%%%%%%%%%%%%%%%%%%%%%%%%%%%%%%%%%%%%%%%%%
None of the  $w^{SS}_\pm({\mathbf p},\l)$ and
$w^{VS}_\pm({\mathbf p},\l)$ four-vector spinors from above is an eigenstate to the $F$
Casimir invariant of the Lorentz algebra in \eqref{FCasm}.
However, the commutation of the spin-$1/2$  projector
$\mathbb{P}^{(1/2)}$ in \eqref{stp}
with the Lorentz projectors $\mathcal{P}^{(j_1)}_F$ in \eqref{pf0},
permits their diagonalizing in the same basis
and allows us to construct $\mathbb{P}^{(1/2)}$ eigenstates that
simultaneously transform according to one of the two $(1/2,j_1)\oplus(j_1,1/2)$ sectors $(j_1=1,0)$
in $\psi_\mu$. These are the states $[w(\mathbf{p},\l)^{(m,1/2)}_{(j_1)}]^\a$ from \eqref{eomsj1}. In the following  however we will need also the quantum number of parity, $\pm$, as a label of the states. For the sake of simplifying notation from now onward we will drop the $(m,1/2)$ label and re-denote the above states by  $[w^{(j_1)}_\pm(\mathbf{p},\l)]^\a$.
\aeq
[w^{(j_1)}_\pm(\mathbf{p},\l)]^\a=N
[\mathcal{P}^{(j_1)}_F]^{\a\b}[w^{SS/SV}_\pm(\mathbf{p},\l)]_\b,
\ceq
where $[w^{SS/SV}_\pm(\mathbf{p},\l)]_\b$ is either a scalar- or a vector-spinor and
$N$ is a normalization factor.

We can now benefit from our knowledge on $w_\pm^{SS}({\mathbf p},\l)$ in \eqref{ssdef} and
$w_\pm^{VS}({\mathbf p},\l)$ in \eqref{vsdef} and  find the  following $j_1=0$ projections:
\aec
~[\mathcal{P}^{(0)}_F]^{\a\b}[w_\pm^{SS}(\mathbf{p},\l)]_\b
&=&\frac{1}{4 m}\g^\a \not{p} u_{\pm}(\mathbf{p},\l),\\
~[\mathcal{P}^{(0)}_F]^{\a\b}[w_\pm^{VS}(\mathbf{p},\l)]_\b
&=&\frac{\sqrt{3}}{4 m}\g^\a \not{p}u_{\pm}(\mathbf{p},\l).
\cec
In this manner the $(1/2,0)\oplus (0,1/2)$ components of
$\omega^{SS}({\mathbf p},\l)$ and $\omega ^{VS}({\mathbf p},\l)$
are identified. In a way similar,
their  $(1/2,1)\oplus (1,1/2)$ components are identified as
\aec
~[\mathcal{P}^{(1)}_F]^{\a\b}[w_\pm^{SS}(\mathbf{p},\l)]_\b
&=&\frac{1}{m}\(p^\a-\frac{1}{4}\not{p}\g^\a\)u_{\pm}(\mathbf{p},\l),\\
~[\mathcal{P}^{(1)}_F]^{\a\b}[w_\pm^{VS}(\mathbf{p},\l)]_\b
&=&-\frac{1}{\sqrt{3}m}\(p^\a-\frac{1}{4}\not{p}\g^\a\)u_{\pm}(\mathbf{p},\l).
\cec

Taking care of the proper normalizations, and in terms of the $[f^{(j_1)}(p)]^\a$ matrices from \eqref{ef0} and \eqref{ef1},
the eigenstates $[w_\pm^{(j_1)}(\mathbf{p},\l)]^\a$ to the combined Lorentz- and Poincar\'e
invariant projectors can finally be cast into the following forms:
\aeq
~[w_\pm^{(j_1)}(\mathbf{p},\l)]^\a=[f^{(j_1)}(\mathbf{p})]^\a
u_{\pm}(\mathbf{p},\l)\label{deflf}.
\ceq
This is
\aec
~[w_\pm^{(0)}(\mathbf{p},\l)]^\a&=&\frac{1}{2 m}\g^\a \not{p}
u_{\pm}(\mathbf{p},\l)\label{defl0s},\\
~[w_\pm^{(1)}(\mathbf{p},\l)]^\a&=&\frac{2}{\sqrt{3}
m}\(p^\a-\frac{1}{4}\g^\a\not{p}\)u_{\pm}(\mathbf{p},\l)\label{defl1s}.
\cec
Our first observation concerns the simplicity of the $j_1=0$ solutions. Next we realize
that the $j_1=1$ solutions obey same relationship,
\aeq
\g^\a[w_\pm^{(1)}(\mathbf{p},\l)]_\a=0,\label{s1aux}
\ceq
as the one already reported  in \eqref{auxj11}. There are two polarizations available for
each parity and two possible parities for each $j_1$ value, making a total of
eight spin-$1/2$ independent states which are now classified according to two distinct irreducible
Lorentz invariant representation spaces.
%%%%%%%%%%%%%%%%%%%%%%%%%%%%%%%%%%%%%%%%%%%%%%%%%%%%%%%%%%%%%%%%%%%%%%%%%%%%%%%%%%%%%%%%%%%%%%%%%%%
\subsubsection{Parity projectors from the  states }\label{sec3b1}
%%%%%%%%%%%%%%%%%%%%%%%%%%%%%%%%%%%%%%%%%%%%%%%%%%%%%%%%%%%%%%%%%%%%%%%%%%%%%%%%%%%%%%%%%%%%%%%%%%%
The  eigenstates of the combined Lorentz-and Poincar\'e invariant projector in \eqref{defl0s}-\eqref{defl1s}
are also of well defined parities, and are normalized  according to
\eqref{deflf} as,
\aeq
~[w_\pm^{(j_1)}(\mathbf{p},\l)]^\a[\overline{w}_\pm^{(j_1)}(\mathbf{p},\l)]_\a=\pm
1.
\ceq
Their  conjugates  are defined as,
\aeq
\overline{w}_\pm^{(j_1)}(\mathbf{p},\l)=[\g^0\overline{w}_\pm^{(j_1)}(\mathbf{p},\l)]^\dagger.
\ceq
Then the parity projectors constructed from \eqref{ppdef} emerge as,
\aeq
~\[\mathbb{P}^{(j_1)}_{\pm}(\mathbf{p})\]_{\a\b}=[f^{(j_1)}(\mathbf{p})]_\a\frac{(\pm\not{p}+m)}{2m}[\overline{f}^{(j_1)}(\mathbf{p})]_\b,
\ceq
meaning that
\aec
~\[\mathbb{P}^{(0)}_{\pm}(\mathbf{p})\]_{\a\b}&=&\frac{1}{2m}\frac{1}{4
m^2}\g_\a \not{p}(\pm\not{p}+m)\not{p}\g_\b,\\
~\[\mathbb{P}^{(1)}_{\pm}(\mathbf{p})\]_{\a\b}&=&\frac{1}{2m}\frac{4}{3
m^2}\(p_\a-\frac{1}{4}\g_\a
\not{p}\)(\pm\not{p}+m)\(p_\b-\frac{1}{4}\not{p}\g_\b\).
\cec
%%%%%%%%%%%%%%%%%%%%%%%%%%%%%%%%%%%%%%%%%%%%%%%%%%%%%%%%%%%%%%%%%%%%%%%%%%%%%%%%%%%%%%%%%%%%%%%%%%%
%%%%%%%%%%%%%%%%%%%%%%%%%%%%%%%%%%%%%%%%%%%%%%%%%%%%%%%%%%%%%%%%%%%%%%%%%%%%%%%%%%%%%%%%%%%%%%%%%%%
%%%%%%%%%%%%%%%%%%%%%%%%%%%%%%%%%%%%%%%%%%%%%%%%%%%%%%%%%%%%%%%%%%%%%%%%%%%%%%%%%%%%%%%%%%%%%%%%%%%
\section{Electromagnetic interaction}\label{sec4}
%%%%%%%%%%%%%%%%%%%%%%%%%%%%%%%%%%%%%%%%%%%%%%%%%%%%%%%%%%%%%%%%%%%%%%%%%%%%%%%%%%%%%%%%%%%%%%%%%%%
Having appropriately constructed the free equations of motion in \eqref{eomsj1}
does not necessarily guarantee that they describe physically observable
particles. Also if we expect to describe electromagnetically interacting particles,
then our equations of motion have also to remain consistent upon the electromagnetic gauging.
The gauging procedure is very sensitive to the momentum dependence of
the wave  equations, especially when they are of second order as are ours.
For this reason before proceeding further we have to attend several details.

We begin with casting the free equations of motion \eqref{eomsj1} for spin-$1/2$
transforming in $(1/2,j_1)\oplus(j_1,1/2)$ as
\aeq
\([\G^{(j_1)}_{\m\n}]_{\a\b}p^\m p^\n-m^2
g_{\a\b}\)[w^{(j_1)}]^\b=0\label{eoms}.
\ceq
Here
\aec
~[\G^{(j_1)}_{\m\n}]_{\a\b}p^\m
p^\n&=&[\mathcal{P}_F^{(j_1)}]_\a{}^\g[\mathcal{T}^{(1/2)}_{\m\n}]_{\g\b}p^\m
p^\nu,\\
~[\mathcal{T}^{(1/2)}_{\m\n}]_{\a\b}p^\m
p^\n&=&m^2[\mathbb{P}^{(1/2)}(\mathbf{p})]_{\a\b}.
\cec
There is certain ambiguity regarding the momentum dependence of the $\mathcal{T}^{(1/2)}_{\m\n}$ tensor
in so far its antisymmetric part is not uniquely fixed within the method.
Due to  the commutativity of the four-momenta $p^\m$ and $p^\n$, it is obvious,
that for free particles  contributions of the type
 $[ \mathcal{T}^{(1/2)} _{\m\n}]_{\g\b}\left[ p^\m p^\n\right]$  nullify.
However, upon gauging,  $p^\mu\rightarrow \pi^\mu=p^\mu-e A^\mu$,
the commutator between the gauged momenta gives rise to the electromagnetic field tensor,
\aeq
[\pi^\m,\pi ^\n]=-i e F^{\m\n}.
\ceq
We here require the antisymmetric part of $\mathcal{T}^{(1/2)}_{\m\n}$ to coincide with
the one emerging from the Lorentz- and Poincar\'e covariant projector as
constructed from the states in \eqref{defl0s}-\eqref{defl1s}.
In so doing, the $\G^{(j_1)}_{\m\n}$ tensor is found as,
\aeq
~[\G^{(j_1)}_{\m\n}]_{\a\b}=[\mathcal{P}_F^{(j_1)}]_\a{}^\g\(\frac{1}{3}\s_{\g\m}\s_{\b\n}+g_{\g\m}g_{\b\n}\),
\ceq
where use of \eqref{stp} has been made. Explicitly for each $j_1$-value we have,
\aec
~[\G^{(0)}_{\m\n}]_{\a\b}&=&\frac{1}{4}\g_\a \g_\m \g_\n \g_\b,\label{defg0}\\
~[\G^{(1)}_{\m\n}]_{\a\b}&=&\frac{4}{3}\(g_{\a\m}-\frac{1}{4}\g_\a
\g_\m\)\(g_{\n\b}-\frac{1}{4}\g_\n \g_\b\),\label{defg1}
\cec
and in reference to \eqref{pf0} and \eqref{pf1}. With
these definitions, both equations of motion  in
\eqref{eoms}  will be shown in the subsequent section to pass the causality test,
and thus qualify  for the description of
electromagnetically interacting spin-$1/2$ particles transforming as
$(1/2,j_1)\oplus(j_1,1/2)$.
%%%%%%%%%%%%%%%%%%%%%%%%%%%%%%%%%%%%%%%%%%%%%%%%%%%%%%%%%%%%%%%%%%%%%%%%%%%%%%%%%%%%%%%%%%%%%%%%%%%
%%%%%%%%%%%%%%%%%%%%%%%%%%%%%%%%%%%%%%%%%%%%%%%%%%%%%%%%%%%%%%%%%%%%%%%%%%%%%%%%%%%%%%%%%%%%%%%%%%%
\subsection{The causality test}\label{sec4a}
%%%%%%%%%%%%%%%%%%%%%%%%%%%%%%%%%%%%%%%%%%%%%%%%%%%%%%%%%%%%%%%%%%%%%%%%%%%%%%%%%%%%%%%%%%%%%%%%%%%
The hyperbolicity and causality of the equations of motion of order $\leq 2$
in the derivatives can be tested using the
Courant-Hilbert method, which requires us to calculate the characteristic
determinant of the gauged equations. In
order to obtain the gauged equations, we first switch in \eqref{eoms} from momentum to position space
using  $[\y^{(j_1)}]^\a=[w^{(j_1)}({\mathbf p},\lambda )]^\a e^{-i x\cdot p}$ as
\aeq
\([\G^{(j_1)}_{\m\n}]_{\a\b}\pd^\m \pd^\n+m^2
g_{\a\b}\)[\y^{(j_1)}]^\b=0.\label{fdeqs}
\ceq
This  equation is in reality a  $16\times 16$ dimensional matrix equation for the
16-component state $[\y^{(j_1)}]^\b$.
However,  considering  only the relevant degrees of freedom of a
spin-$1/2$ particle (regardless of its
parity) we have to arrange the above equation as a $4\times 4$ matrix equation
acting on a 4-component state vector  as indicated by
the explicit form of the solutions in terms of
$4$-component spinors. Then, according to the
gauge principle,  we couple this equations minimally to  an electromagnetic field
according to
\aeq
\pd\rightarrow D=\pd+i e A,
\ceq
where $e$ is the electric charge of the particle. The characteristic
determinant is then found by replacing the
highest order  derivatives  by the components of
the vector $n^\m$, normal to the characteristic surfaces,  and which
characterizes  the propagation of the (classical)  wave fronts of the gauged equation.
If the vanishing  of the
characteristic determinant demands to have a real-valued time-like component
$n^0$, then the equation
is hyperbolic. If this determinant nullifies as $n^\m n_\m=0$, then the
equation is in addition  causal \cite{Velo:1970ur}.
%%%%%%%%%%%%%%%%%%%%%%%%%%%%%%%%%%%%%%%%%%%%%%%%%%%%%%%%%%%%%%%%%%%%%%%%%%%%%%%%%%%%%%%%%%%%%%%%%%%%%%%%%%%%%%%%%%%%%%%
\subsubsection{Gauging the wave  equation for the
 {$(1/2,0)\oplus(0,1/2)$} sector}\label{sec4a1}
%%%%%%%%%%%%%%%%%%%%%%%%%%%%%%%%%%%%%%%%%%%%%%%%%%%%%%%%%%%%%%%%%%%%%%%%%%%%%%%%%%%%%%%%%%%%%%%%%%%%%%%%%%%%%%%%%%%%%%%
In order to find the explicit form of the gauged equation for the case under consideration,
we first substitute  $\G^{(0)}_{\m\n}$ in \eqref{fdeqs} by its definition in  \eqref{defg0},
and then, in making use of  \eqref{defl0s}, and  $\y=u_\pm({\mathbf p},\l)  e^{-ip\cdot x}$, we arrive at
\aeq
\(\frac{1}{4}\g_\a \g_\m \g_\n \g_\b \pd^\m \pd^\n +m^2 g_{\a\b}\)\(\frac{-i}{2 m}\g^\b
\not{\pd}\)\y=0.
\ceq
This equation arranges to  a
$4\times4$ matrix equation upon contraction by $\g^\a$ from the left, then 
the factorization of a $\g_\b$-matrix  becomes possible, with the result,
\aeq
\(\g_\m \g_\n \pd^\m \pd^\n +m^2\)\g_\b\(\frac{-i}{2 m}\g^\b
\not{\pd}\)\y=0.\label{j0cfact}
\ceq
Carrying out now the $\gamma_\beta\gamma^\beta$ contraction amounts to,
\aeq
\frac{2}{m}\(\g_\m \g_\n \pd^\m \pd^\n+m^2\)\(-i\not{\pd}\)\y=0,
\ceq
which leads to the following  gauged equation ,
\aeq
\frac{2}{m}\(\g_\m \g_\n D^\m D^\n+m^2\)\(-i\not{D}\)\y=0.
\label{Geq0}
\ceq
This equation is of third order in derivatives,
but it favorably factorizes into  a quadratic and a linear equation,
thus allowing us to apply the  Courant-Hilbert criterion to each one of
the factors separately \cite{Rico:2007br}. In so doing, the characteristic determinant
also factorizes into two characteristic determinants of the respective
quadratic and linear equations.
In this fashion,  it becomes possible to test the causality of the wave equation for the $j_1=0$.
We calculate
\aeq
\mathcal{D}^{(0)}(n)=\mathcal{D}^{(0)}_1(n) \mathcal{D}^{(0)}_2(n)=\(\frac{2}{m}\)^4(n^2)^4
(n^2)^2\label{det0},
\ceq
where
\aec
\mathcal{D}_1^{(0)}(n)&=&\left\vert -\frac{2}{m}\g_\m \g_\n n^\m n^\n\right\vert=\left\vert -\frac{2}{m}n^2 \right\vert=
\(\frac{2}{m}\)^4(n^2)^4,\label{detj01}\\
\mathcal{D}_2^{(0)}(n)&=&\vert \not{n}\vert=(n^2)^2\label{detj02}.
\cec
Nullifying $D^{(0)}(n)$ in \eqref{det0} amounts to the
condition $n^2=n^\m n_\m=0$, which is the accepted  indicator of causal propagation.
%%%%%%%%%%%%%%%%%%%%%%%%%%%%%%%%%%%%%%%%%%%%%%%%%%%%%%%%%%%%%%%%%%%%%%%%%%%%%%%%%%%%%%%%%%%%%%%%%%%%%%%%%%%%%%%%%%%%%%%
\subsubsection{Gauging the wave equation for the
 {$(1/2,1)\oplus(1,1/2)$} sector}\label{sec4a2}
%%%%%%%%%%%%%%%%%%%%%%%%%%%%%%%%%%%%%%%%%%%%%%%%%%%%%%%%%%%%%%%%%%%%%%%%%%%%%%%%%%%%%%%%%%%%%%%%%%%%%%%%%%%%%%%%%%%%%%%
Using  \eqref{defg1} to substitute for $\G^{(1)}_{\m\n}$ in \eqref{fdeqs} together with the
explicit forms of the \eqref{fdeqs}  solutions  from
\eqref{defl1s} amounts to,
\aeq
\frac{4}{3}\[\(g_{\a\m}-\frac{1}{4}\g_\a \g_\m\)\(g_{\n\b}-\frac{1}{4}\g_\n
\g_\b\)\pd^\m \pd^\n+m^2g_{\a\b}\]
\frac{2}{\sqrt{3} m}\(-i\pd^\b+i\frac{1}{4}\g^\b \not{\pd}\)\y=0,
\label{Gl1}
\ceq
with $\y=u_\pm({\mathbf p},\l) e^{-ix\cdot p}$. Then by virtue of the auxiliary condition
\eqref{s1aux} the equation \eqref{Gl1} simplifies as,
\aeq
\frac{4}{3}\[\(g_{\a\m}-\frac{1}{4}\g_\a \g_\m\)\pd^\m \pd_\b+m^2g_{\a\b}\]
\frac{-2i}{\sqrt{3} m}\(\pd^\b-\frac{1}{4}\g^\b \not{\pd}\)\y=0.
\ceq
In order to obtain a $ 4\times 4$ matrix equation
we perform a contraction by   $\pd^\a$ arriving at,
\aeq
\frac{4}{3}\[\(\pd_\m-\frac{1}{4}\not{\pd} \g_\m\)\pd^\m\pd^\b
+m^2\pd^\b\]\label{j1cfact}
\frac{-2i}{\sqrt{3} m}\(\pd^\b-\frac{1}{4}\g^\b \not{\pd}\)\y=0.
\ceq
This equation can then be factorized into two quadratic equations,
\aeq
\frac{4}{3}\frac{(-2i)}{\sqrt{3} m}\[\(\pd_\m-\frac{1}{4}\not{\pd} \g_\m\)\pd^\m
+m^2\]\[\pd_\b
\(\pd^\b-\frac{1}{4}\g^\b \not{\pd}\)\]\y=0.
\ceq
Notice that we have not made any use of the commutativity of the $\pd$-derivatives,
so that the gauged equation can be written as
\aeq
\frac{4}{3}\frac{(-2i)}{\sqrt{3} m}\[\(D_\m-\frac{1}{4}\not{D} \g_\m\)D^\m
+m^2\]\[D_\b
\(D^\b-\frac{1}{4}\g^\b \not{D}\)\]\y=0.
\ceq
The characteristic determinant for $j_1=1$ is then found as the
product of the following two determinants:
\aeq\label{det1}
\mathcal{D}^{(1)}(n)=\(\frac{4}{3}\frac{(-2i)}{\sqrt{3}
m}\)^4\mathcal{D}^{(1)}_1(n) \mathcal{D}^{(1)}_2(n)
=\(\frac{4}{3}\frac{2}{\sqrt{3} m}\)^4\(\frac{3}{4}\)^8 (n^2)^4
(n^2)^4=\(\frac{\sqrt{3}}{2m}\)^4 (n^2)^4 (n^2)^4,
\ceq
where
\aec
\mathcal{D}_1^{(1)}(n)&=&\left\vert \(-n_\m+\frac{1}{4}\not{n}
\g_\m\)n^\m\right\vert=\left\vert -\frac{3}{4}n^2
\right\vert=\(\frac{3}{4}\)^4 (n^2)^4,\\
\mathcal{D}_2^{(1)}(n)&=&\left\vert n_\b
\(-n^\b+\frac{1}{4}\g^\b\not{n}\)\right\vert=\left\vert -\frac{3}{4}n^2
\right\vert=\(\frac{3}{4}\)^4 (n^2)^4.
\cec
Again, the condition for the determinant \eqref{det1} to be zero, holds valid for
$n^2=n^\m n_\m=0$, thus ensuring causality. This result
completes the proof of the causality of the equations of motion for both
spin-$1/2$ sectors in $\psi_\mu$, transforming in $(1/2,j_1)\oplus(j_1,1/2)$ with $j_1=0,1$.
%%%%%%%%%%%%%%%%%%%%%%%%%%%%%%%%%%%%%%%%%%%%%%%%%%%%%%%%%%%%%%%%%%%%%%%%%%%%%%%%%%%%%%%%%%%%%%%%%%%%%%%%%%%%%%%%%%%%%%%
%%%%%%%%%%%%%%%%%%%%%%%%%%%%%%%%%%%%%%%%%%%%%%%%%%%%%%%%%%%%%%%%%%%%%%%%%%%%%%%%%%%%%%%%%%%%%%%%%%%%%%%%%%%%%%%%%%%%%%%
\subsection{The Lagrangians for the lower spin-$1/2$ sectors of the Rarita-Schwinger four-vector field}\label{sec4b}
\begin{figure}
\includegraphics{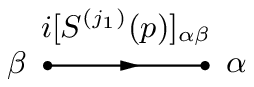}
\caption{\label{propfig} Feynman rule for the propagators of particles in the
$(1/2,j_1)\oplus(j_1,1/2)$ sector of the four-vector-spinor. They are obtained
as the inverse of the equations of motion, the explicit form of $S^{(j_1)}(p)$
for each $j_1$-value is given in \eqref{props0}-\eqref{deltaj1}.}

\includegraphics{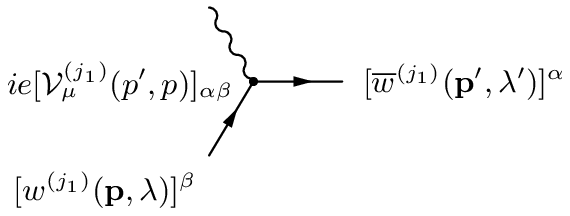}
\caption{\label{regla1fig} Feynman rule for the one-photon vertex with
$(1/2,j_1)\oplus(j_1,1/2)$ particles. It comes from the interaction Lagrangian
\eqref{lint}, for the specific definition of $\mathcal{V}^{(j_1)}_\m(p',p)$
corresponding to each $j_1$-value see \eqref{vmdef}.}

\includegraphics{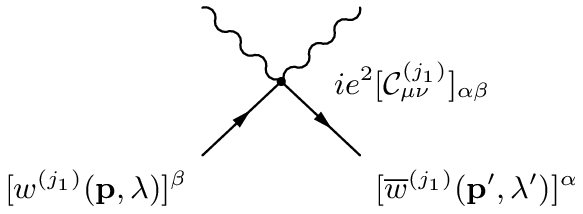}
\caption{\label{regla2fig} Diagram for the two-photons Feynman rule
corresponding to the interaction Lagrangian \eqref{lint}, the explicit form of
the $\mathcal{C}^{(j_1)}_{\m\n}$ vertex is given in \eqref{cmndef}.}
\end{figure}

%%%%%%%%%%%%%%%%%%%%%%%%%%%%%%%%%%%%%%%%%%%%%%%%%%%%%%%%%%%%%%%%%%%%%%%%%%%%%%%%%%%%%%%%%%%%%%%%%%%%%%%%%%%%%%%%%%%%%%%
For practical calculations it is advantageous  to have at ones disposal a  gauged Lagrangian,
out of which one can deduce  the Feynman rules of the theory. The Lagrangians for positive and negative parity states 
usually differ only by an overall sign, it compensates for the different normalizations of states of opposite parities.
With  this in mind and for the sake of concreteness, in the following we shall only deal with Lagrangians written in 
terms of the positive parity states. The free equations of motion \eqref{fdeqs} for the positive parity states relate 
to the Lagrangian by the Euler-Lagrange equations. For second order equations of motion the corresponding Lagrangians 
are of the form,
\aeq
\mathcal{L}^{(j_1)}_{\text{free}}=(\pd^\m[\overline{\y}^{(j_1)}]^\a)
[\G^{(j_1)}_{\m\n}]_{\a\b} \pd^\n [\y^{(j_1)}]^\b
-m^2[\overline{\y}^{(j_1)}]^\a[\y^{(j_1)}]_\a,
\ceq
and the gauged Lagrangian is obtained as usually by the replacement of the
ordinary-  by covariant derivatives,
\aeq
\mathcal{L}^{(j_1)}=(D^{\m*}[\overline{\y}^{(j_1)}]^\a)
[\G^{(j_1)}_{\m\n}]_{\a\b} D^\n
[\y^{(j_1)}]^\b-m^2[\overline{\y}^{(j_1)}]^\a[\y^{(j_1)}]_\a.
\ceq
Now  $\mathcal{L}^{(j_1)}$  can be decomposed  into free and interaction Lagrangians as,
\aec
\mathcal{L}^{(j_1)}&=&\mathcal{L}^{(j_1)}_{\text{free}}+\mathcal{L}^{(j_1)}_{\text{int}},\\
\mathcal{L}^{(j_1)}_{\text{int}}&=&-j_\m^{(j_1)}A^\m+k^{(j_1)}_{\m\n} A^\m
A^\n\label{lint}.
\cec
Here, $j_\m^{(j_1)}$ is the electromagnetic current, while $k^{(j_1)}_{\m\n}$ is
the structure of a two-photon coupling. In momentum space, and for the positive parity states $[w^{(j_1)}({\mathbf p},\l )]^\b$ we find,
\aec
j_\m^{(j_1)}&=& e
[\overline{w}^{(j_1)}(\mathbf{p}', \l^\prime )]^\a[\mathcal{V}_{\m}^{(j_1)}(p',p)]_{\a\b}[w^{(j_1)}(\mathbf{p},\l )]^\b,\label{currentj1}\\
k_{\m\nu}^{(j_1)}&=&e^2[\overline{w}^{(j_1)}(\mathbf{p}',\l^\prime )]^\a[\mathcal{C}_{\m\n}^{(j_1)}]_{\a\b}[w^{(j_1)}(\mathbf{p},\l )]^\b,
\cec
where $[\mathcal{V}_{\m}^{(j_1)}(p',p)]_{\a\b}$ and
$[\mathcal{C}_{\m\n}^{(j_1)}]_{\a\b}$ are the one- and two-photon
vertexes which, together with the propagators \eqref{props}, determine the
Feynman rules. The latter are  depicted on the Figs. \ref{propfig},\ref{regla1fig},\ref{regla2fig}.
\aec
~[\mathcal{V}_{\m}^{(j_1)}(p',p)]_{\a\b}&=&[\G^{(j_1)}_{\n\m}]_{\a\b}p'^{\n}+[\G^{(j_1)}_{\m\n}]_{\a\b}p^{\n},\label{vmdef}\\
~[\mathcal{C}_{\m\n}^{(j_1)}]_{\a\b}&=&\frac{1}{2}\(
[\G^{(j_1)}_{\m\n}]_{\a\b}+[\G^{(j_1)}_{\n\m}]_{\a\b}\)\label{cmndef}.
\cec
%---------------------------Figures------------------------------------------------------------------------------------
%^^^^^^^^^^^^^^^^^^^^^^^^^^^^^^^^^^^^^^^^^^^^^^^^^^^^^^^^^^^^^^^^^^^^^^^^^^^^^^^^^^^^^^^^^^^^^^^^^^^^^^^^^^^^^^^^^^^^^^
%---------------------------Figures------------------------------------------------------------------------------------
%^^^^^^^^^^^^^^^^^^^^^^^^^^^^^^^^^^^^^^^^^^^^^^^^^^^^^^^^^^^^^^^^^^^^^^^^^^^^^^^^^^^^^^^^^^^^^^^^^^^^^^^^^^^^^^^^^^^^^^
In particular, $[\mathcal{V}_{\m}^{(j_1)}(p',p)]_{\a\b}$ obeys the
Ward-Takahashi identity,
\aeq
(p'-p)^\m
[\mathcal{V}_{\m}^{(j_1)}(p',p)]_{\a\b}=[S^{(j_1)}(p')]_{\a\b}^{-1}-[S^{(j_1)}(p)]_{\a\b}^{-1}\label{wti},
\ceq
where $[S^{(j_1)}(p)]_{\a\b}$ are the propagators in \eqref{props}.
This relationship  leads to gauge invariance of the
amplitudes which define the Compton scattering process.
However, before evaluating  this
process it is very instructive to figure
out the values of the magnetic dipole moments of the particles under consideration
prescribed by the currents in \eqref{currentj1}.
%%%%%%%%%%%%%%%%%%%%%%%%%%%%%%%%%%%%%%%%%%%%%%%%%%%%%%%%%%%%%%%%%%%%%%%%%%%%%%%%%%%%%%%%%%%%%%%%%%%%%%%%%%%%%%%%%%%%%%%
%%%%%%%%%%%%%%%%%%%%%%%%%%%%%%%%%%%%%%%%%%%%%%%%%%%%%%%%%%%%%%%%%%%%%%%%%%%%%%%%%%%%%%%%%%%%%%%%%%%%%%%%%%%%%%%%%%%%%%%
\subsection{Magnetic dipole moments}\label{sec4c}
%%%%%%%%%%%%%%%%%%%%%%%%%%%%%%%%%%%%%%%%%%%%%%%%%%%%%%%%%%%%%%%%%%%%%%%%%%%%%%%%%%%%%%%%%%%%%%%%%%%%%%%%%%%%%%%%%%%%%%%
We begin with the currents in  momentum space in \eqref{currentj1} and  for positive
parity states with polarization $\l$, which are given  in terms of Dirac's $u$-spinors by,
\aeq\label{j1ov}
j_\m^{(j_1)}(\mathbf{p},\mathbf{p'})= e \overline{u}(\mathbf{p}',\l)
\widetilde{\mathcal{V}}^{(j_1)}_\m({p}',{p}) u(\mathbf{p},\l).
\ceq
Here we have used \eqref{deflf} and found,
\aec
\widetilde{\mathcal{V}}^{(j_1)}_\m({p'},{p})&=&[\overline{f}^{(j_1)}(\mathbf{p}')]^\a
[\mathcal{V}_{\m}^{(j_1)}(p',p)]_{\a\b}[f^{(j_1)}(\mathbf{p})]^\b.
\cec
Incorporation of the  mass-shell condition amounts to,
\aec
j_\m^{(0)}(\mathbf{p'},\mathbf{p})&=& e\, \overline{u}(\mathbf{p}',\l)\(
2m\g_\m \)u(\mathbf{p},\l),\label{j0gd}\\
j_\m^{(1)}(\mathbf{p'},\mathbf{p})&=& e\, \overline{u}(\mathbf{p}',\l)
\(\frac{4}{3}(p'+p)_\m-\frac{2m}{3}\g_\m \) u(\mathbf{p},\l)\label{j1gd}.
\cec
One immediately notices that for $j_1=0$, the textbook Dirac
current is recovered, as it should be and in accord with the reducibility of the the four-vector spinor
discussed in the introduction.
We now make use of the Gordon decomposition of the Dirac current,
\aeq
2m\,e\, \overline{u}(\mathbf{p}',\l)\g_\m
u(\mathbf{p},\l)=e\,\overline{u}(\mathbf{p}',\l)\[ (p'+p)_\m+2 i
M^S_{\m\n}(p'-p)^\n \]u(\mathbf{p},\l).
\ceq
Here $M^S_{\m\n}$ are the elements of the Lorentz algebra in $(1/2,0)\oplus (0,1/2)$
in \eqref{genss}, while the factor 2 in front of them stands for the gyromagnetic ratio. As a result, the currents in \eqref{j0gd} and \eqref{j1gd} take the form
\aeq
j_\m^{(j_1)}(\mathbf{p'},\mathbf{p})=e\,\overline{u}(\mathbf{p}',\l)\[
(p'+p)_\m+i g^{(j_1)}M^S_{\m\n}(p'-p)^\n \]u(\mathbf{p},\l),
\ceq
where
\aec
g^{(0)}&=&2\label{g0},\\
g^{(1)}&=&-\frac{2}{3}\label{g1}.
\cec
The above eqs.~\eqref{g0} and \eqref{g1} show that the electromagnetic currents for particles transforming in
$(1/2,j_1)\oplus(j_1,1/2)$ are characterized by different magnetic dipole moments for
different $j_1$ values. The gauged
Lagrangian corresponding to the combined  Lorentz- and Poincar\'e invariant  projector,
that describes particles of charge $e$  transforming in
$(1/2,0)\oplus(0,1/2)$ predicts the following magnetic moment,
\aeq
\m^{(0)}(\l)=2\frac{\l e}{2 m}\label{dipmm0}.
\ceq
The latter coincides with the standard value for a Dirac particle of polarization $\l$.
Instead,  the Lagrangian of same type  predicts for  particles of charge $e$
transforming in  $(1/2,1)\oplus(1,1/2)$  a magnetic dipole moment of
\aeq
\m^{(1)}(\l)=-\frac{2}{3}\frac{\l e}{2 m}\label{dipmm1}.
\ceq
This dependence of the magnetic dipole moment on the space-time transformation properties
of the particle, is similar to the one
found for high-spin states, where particles with equal spins, transforming in
different representation spaces  of the Lorentz algebra,
have also been observed to be characterized by different sets of electromagnetic multipole moments
\cite{DelgadoAcosta:2012yc}. An
electromagnetic process however is not entirely determined by the
electromagnetic multipole moments of the particles,
which by definition are associated with the on-shell states.
It is basically determined  by the
complete gauged Lagrangian. Remarkable,  different Lagrangians can lead to the same multipole moments
\cite{Lorce:2009bs}, \cite{DelgadoAcosta:2012yc}. The
knowledge on the  electromagnetic multipole  moments
is therefore  not sufficient to completely characterize a theory.
The more profound test for a Lagrangian regards processes
involving off-shell  states. One such process, the Compton scattering,
is considered in detail the next section.
%%%%%%%%%%%%%%%%%%%%%%%%%%%%%%%%%%%%%%%%%%%%%%%%%%%%%%%%%%%%%%%%%%%%%%%%%%%%%%%%%%%%%%%%%%%%%%%%%%%%%%%%%%%%%%%%%%%%%%%
%%%%%%%%%%%%%%%%%%%%%%%%%%%%%%%%%%%%%%%%%%%%%%%%%%%%%%%%%%%%%%%%%%%%%%%%%%%%%%%%%%%%%%%%%%%%%%%%%%%%%%%%%%%%%%%%%%%%%%%
%%%%%%%%%%%%%%%%%%%%%%%%%%%%%%%%%%%%%%%%%%%%%%%%%%%%%%%%%%%%%%%%%%%%%%%%%%%%%%%%%%%%%%%%%%%%%%%%%%%%%%%%%%%%%%%%%%%%%%%
\section{Compton scattering off spin-$1/2$ in $(1/2,j_1)\oplus(j_1,1/2)$}\label{sec5}
\begin{figure}
\includegraphics{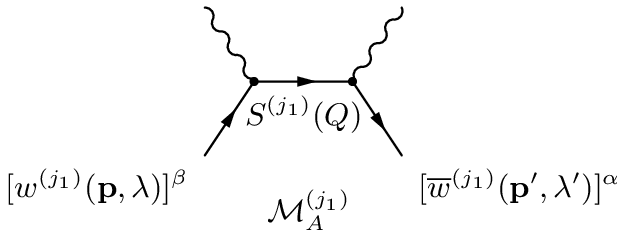}
\caption{\label{mafig} Diagram for the direct-scattering contribution
\eqref{maj1} to the Compton scattering amplitude \eqref{camp}.}

\includegraphics{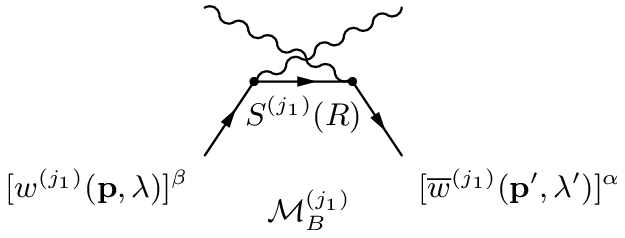}
\caption{\label{mbfig} Diagram for the exchange-scattering contribution
\eqref{mbj1} to the Compton scattering amplitude \eqref{camp}.}

\includegraphics{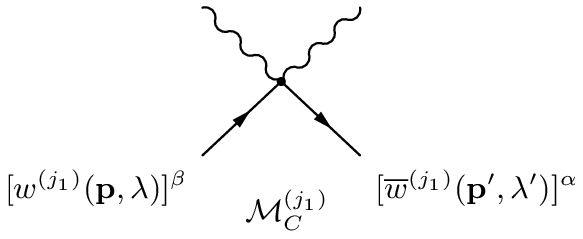}
\caption{\label{mcfig} Diagram for the point-scattering contribution
\eqref{mcj1} to the Compton scattering amplitude \eqref{camp}.}
\end{figure}

%%%%%%%%%%%%%%%%%%%%%%%%%%%%%%%%%%%%%%%%%%%%%%%%%%%%%%%%%%%%%%%%%%%%%%%%%%%%%%%%%%%%%%%%%%%%%%%%%%%%%%%%%%%%%%%%%%%%%%%
The Compton scattering amplitudes are constructed in terms of the Feynman
rules (shown in the Figs. \ref{propfig},\ref{regla1fig},\ref{regla2fig}) for each $j_1$- value, giving
\aeq\label{camp}
\mathcal{M}^{(j_1)}=\mathcal{M}_A^{(j_1)}+\mathcal{M}_B^{(j_1)}+\mathcal{M}_C^{(j_1)},
\ceq
where $\mathcal{M}_A^{(j_1)}$, $\mathcal{M}_B^{(j_1)}$, $\mathcal{M}_C^{(j_1)}$ correspond to the amplitudes
for direct, exchange and point scatterings, respectively. In the following we use the symbols $p$ and $p'$ to denote the momentum of the incident and 
scattered spin-$1/2$ particles respectively, while using $q$ and $q'$ to denote the momentum of the incident and scattered photons respectively, so that
\aec
i\mathcal{M}_A^{(j_1)}&=&e^2[\overline{w}^{(j_1)}(\mathbf{p}',\l')]^{\a}~[U^{(j_1)}_{\m\n}(p',Q,p)]_{\a\b}
[ w(\mathbf{p},\l)^{(j_1)}]^{\b}[\e^\m(\mathbf{q}',\ell')]^* \e^\n(\mathbf{q},\ell),\label{maj1}\\
i\mathcal{M}_B^{(j_1)}&=&e^2[\overline{w}^{(j_1)}(\mathbf{p}',\l')]^{\a}~[U^{(j_1)}_{\n\m}(p',R,p)]_{\a\b}
[ w(\mathbf{p},\l)^{(j_1)}]^{\b}[\e^\m(\mathbf{q}'.\ell')]^* \e^\n(\mathbf{q},\ell),\label{mbj1}\\
i\mathcal{M}_C^{(j_1)}&=&-e^2[\overline{w}^{(j_1)}(\mathbf{p}',\l')]^{\a}~[\mathcal{C}_{\m\n}^{(0)}+\mathcal{C}_{\n\m}^{(0)}]_{\a\b}
[ w(\mathbf{p},\l)^{(j_1)}]^{\b}[\e^\m(\mathbf{q}',\ell')]^* \e^\n(\mathbf{q},\ell)\label{mcj1},
\cec
where $Q=p+p'=q+q'$ and $R=p'-q=p-q'$ stand for the momentum of the intermediate states and
\aeq
~[U^{(j_1)}_{\m\n}(p',Q,p)]_{\a\b}=[\mathcal{V}^{(j_1)}_{\m}(p',Q)]_{\a\g}
[S^{(j_1)}(Q)]^{\g\d}[\mathcal{V}^{(j_1)}_\n(Q,p)]_{\d\b}.
\ceq
These amplitudes are shown in the Figs. \ref{mafig},\ref{mbfig},\ref{mcfig}.
%^^^^^^^^^^^^^^^^^^^^^^^^^^^^^^^^^^^^^^^^^^^^^^^^^^^^^^^^^^^^^^^^^^^^^^^^^^^^^^^^^^^^^^^^^^^^^^^^^^^^^^^^^^^^^^^^^^^^^^
%---------------------------Figures------------------------------------------------------------------------------------
%---------------------------Figures------------------------------------------------------------------------------------
%^^^^^^^^^^^^^^^^^^^^^^^^^^^^^^^^^^^^^^^^^^^^^^^^^^^^^^^^^^^^^^^^^^^^^^^^^^^^^^^^^^^^^^^^^^^^^^^^^^^^^^^^^^^^^^^^^^^^^^
Their gauge invariance  is ensured by the
Ward-Takahashi identity \eqref{wti} (see for example
\cite{DelgadoAcosta:2009ic}). In the following we present the calculation for
each $j_1$-value separately.
%%%%%%%%%%%%%%%%%%%%%%%%%%%%%%%%%%%%%%%%%%%%%%%%%%%%%%%%%%%%%%%%%%%%%%%%%%%%%%%%%%%%%%%%%%%%%%%%%%%%%%%%%%%%%%%%%%%%%%%
%%%%%%%%%%%%%%%%%%%%%%%%%%%%%%%%%%%%%%%%%%%%%%%%%%%%%%%%%%%%%%%%%%%%%%%%%%%%%%%%%%%%%%%%%%%%%%%%%%%%%%%%%%%%%%%%%%%%%%%
\subsection{Compton scattering off particles in the single-spin
 {$(1/2,0)\oplus(0,1/2)$} {} sector of the four-vector--spinor }\label{sec5a}
%%%%%%%%%%%%%%%%%%%%%%%%%%%%%%%%%%%%%%%%%%%%%%%%%%%%%%%%%%%%%%%%%%%%%%%%%%%%%%%%%%%%%%%%%%%%%%%%%%%%%%%%%%%%%%%%%%%%%%%
For the case under investigation it is very useful to write the amplitudes in terms of $u$-spinors:
\aec
i\mathcal{M}_A^{(0)}&=&e^2\overline{u}(\mathbf{p}',\l')~\widetilde{U}^{(0)}_{\m\n}(p',Q,p)u(\mathbf{p},\l)[\e^\m(\mathbf{q}',\ell')]^*
\e^\n(\mathbf{q},\ell),\\
i\mathcal{M}_B^{(0)}&=&e^2\overline{u}(\mathbf{p}',\l')~\widetilde{U}^{(0)}_{\n\m}(p',R,p)u(\mathbf{p},\l)[\e^\m(\mathbf{q}',\ell')]^*
\e^\n(\mathbf{q},\ell),\\
i\mathcal{M}_C^{(0)}&=&-e^2\overline{u}(\mathbf{p}',\l')~\widetilde{\mathcal{C}}^{(0)}_{\m\n}u(\mathbf{p},\l)[\e^\m(\mathbf{q}',\ell')]^*
\e^\n(\mathbf{q},\ell),
\cec
where
\aec
\widetilde{U}^{(j_1)}_{\m\n}(p',Q,p)&=&[\overline{f}^{(j_1)}(\mathbf{p}')]^\a
[U^{(j_1)}_{\m\n}(p',Q,p)]_{\a\b}[f^{(j_1)}(\mathbf{p})]^\b,\label{utrans}\\
\widetilde{\mathcal{C}}^{(j_1)}_{\m\n}&=&[\overline{f}^{(j_1)}(\mathbf{p}')]^\a [\mathcal{C}_{\m\n}^{(j_1)}+\mathcal{C}^{(j_1)}_{\n\m}]_{\a\b}[f^{(j_1)}(\mathbf{p})]^\b\label{ctrans}
\cec
with the $[f^{(0)}(\mathbf{p})]^\a$ matrices taken from \eqref{ef0}. Making
use of the explicit form of the propagator in  \eqref{props}, on the one side,
and the vertices in \eqref{vmdef}, \eqref{cmndef} with $j_1=0$, one the other side,
we  find,
\aec
\widetilde{U}^{(0)}_{\m\n}(p',Q,p)&=&2m \g_\m
\frac{\not{Q}+m}{Q^2-m^2}\g_\n+\g_{\m}\g_{\n},\\
\widetilde{U}^{(0)}_{\n\m}(p',R,p)&=&2m \g_\n
\frac{\not{R}+m}{R^2-m^2}\g_\m+\g_{\n}\g_{\m},\\
\widetilde{\mathcal{C}}^{(0)}_{\m\n}&=&2g_{\m\n}.
\cec
Here,  we achieved some simplifications by replacing by $m$ all appearances of
$\not{p'}$, and $\not{p}$ as well on the left as on the right, respectively,
in noticing that  these always act on $u$-spinors of positive parity only.
The complete amplitude then emerges as,
\aeq
i\mathcal{M}^{(0)}=2m e^2\overline{u}(\mathbf{p}',\l')\(\g_\m
\frac{\not{Q}+m}{Q^2-m^2}\g_\n+\g_\n
\frac{\not{R}+m}{R^2-m^2}\g_\m\)u(\mathbf{p},\l)[\e^\m(\mathbf{q}',\ell')]^* \e^\n(\mathbf{q},\ell),
\ceq
which is just the Compton scattering amplitude associated with the Dirac
Lagrangian for states normalized to unity. This expression  can be further extended
toward  an arbitrary magnetic dipole moment, a subject of the next section.
%%%%%%%%%%%%%%%%%%%%%%%%%%%%%%%%%%%%%%%%%%%%%%%%%%%%%%%%%%%%%%%%%%%%%%%%%%%%%%%%%%%%%%%%%%%%%%%%%%%%%%%%%%%%%%%%%%%%%%%
\subsubsection{Allowing for an arbitrary  {$g$}-factor for
 {$(1/2,0)\oplus(0,1/2)$} particles}\label{sec5a1}
%%%%%%%%%%%%%%%%%%%%%%%%%%%%%%%%%%%%%%%%%%%%%%%%%%%%%%%%%%%%%%%%%%%%%%%%%%%%%%%%%%%%%%%%%%%%%%%%%%%%%%%%%%%%%%%%%%%%%%%
The factorization in \eqref{j0cfact} and the calculation of the
characteristic determinant in \eqref{detj01}, together with the freedom of choice of
the antisymmetric part of the $[\G_{\m\n}^{(0)}]_{\a\b}$ tensor admitted by
the Poincar\'e covariant projector method,  allows us to make the following extension:
\aeq
[\G_{\m\n}^{(0)}]_{\a\b}\rightarrow
[\G_{\m\n}^{(0)}(g)]_{\a\b}=[\G_{\m\n}^{(0)}]_{\a\b}+\frac{i}{4}(2-g)\g_\a
M^S_{\m\n} \g_\b\label{j0ext}.
\ceq
The  magnetic dipole moment of a particle with polarization $\l$ corresponding
to this extension is then given by,
\aeq
\m^{(0)}(g,\l)=g\frac{e\l}{2m},
\ceq
thus leading to an arbitrary $g$-factor counterpart to our previous
magnetic dipole moment of the fixed value $g^{(0)}=2$ in \eqref{dipmm0}. The Compton scattering
calculations for this extension requires to incorporate the replacement in  \eqref{j0ext}
into the Feynman rules for $j_1=0$. In the calculation of the squared amplitude we use
the following formulas for any $j_1$:
\aec\label{m2av}
\overline{\left\vert
\mathcal{M}^{(j_1)}\right\vert^2}&=&\frac{1}{4}\sum_{\l,\l',\ell,\ell'}
[\mathcal{M}^{(j_1)}][\mathcal{M}^{(j_1)}]^\dagger\\
&=&Tr\left[(\mathcal{M}^{(j_1)})_{\m\n}(p',Q,R,p)
(\mathcal{M}^{(j_1)})^{\n\m}(p,R,Q,p')\right]
\cec
where we have defined
\aec
\mathcal{M}^{(j_1)}_{\m\n}(p',Q,R,p)&=&\frac{e^2}{2}\(\frac{\not{p}'+m}{2m}\)\mathbb{U}^{(j_1)}_{\m\n}(p',Q,R,p),\\
\mathbb{U}^{(j_1)}_{\m\n}(p',Q,R,p)&=&\widetilde{U}^{(j_1)}_{\m\n}(p',Q,p)+\widetilde{U}^{(j_1)}_{\n\m}(p',R,p)-\widetilde{\mathcal{C}}^{(j_1)}_{\m\n},
\cec
with $\widetilde{U}$ and $\widetilde{\mathcal{C}}$ taken from \eqref{utrans}
and \eqref{ctrans}. Here we have also used \eqref{ppdef} and:
\aeq
~[\mathbb{P}^{(j_1)}_+(\mathbf{p})]_{\a\b}=[f^{(j_1)}(\mathbf{p})]_\a\(\frac{\not{p}+m}{2m}\)[\overline{f}^{(j_1)}(\mathbf{p})]_\b,
\ceq
for the spin-1/2 target particles of positive parity and
\aeq
\sum_{\ell}\e^\m(\mathbf{q},\ell)[\e^\n(\mathbf{q},\ell)]^{*}=-g^{\m\n}\label{phpr},
\ceq
for the polarization vectors of the photons. The result for the averaged
squared amplitude for $j_1=0$ with an arbitrary $g$-factor ends up being
\begin{eqnarray}\label{m2nkr12g}
\overline{\vert{\mathcal M}^{(0)}(g^{(0)})\vert^2}&=& f_0+f_D
+\frac{e^4(2 m^2-s-u)}{16 m^2 \left(m^2-s\right)^2
\left(m^2-u\right)^2}\sum_{k=1}^4 (g^{(0)}-2)^k a_{k},
\end{eqnarray}
where $s,u$ are the standard Mandelstam variables and we are using the notations
\aec
f_0&=& \frac{4e^4(5m^8-4(s+u)m^6+(s^2+u^2)+s^2 u^2)}{(m^2-s)^2(m^2-u)^2},\\
f_D&=&-\frac{2e^4(-2m^2+s+u)^2}{(m^2-s)(m^2-u)}.
\cec
Here, $f_0$ is the Compton scattering squared amplitude corresponding to
spin-0 particles \cite{DelgadoAcosta:2009ic} and $(f_0+f_D)$ is the standard averaged squared amplitude
for Compton scattering coming from the Dirac Lagrangian. We furthermore have defined:
\begin{eqnarray}
a_1&=&-32 m^2 \left(m^2-s\right) \left(m^2-u\right) \left(2 m^2-s-u\right),\label{ca1}\\
a_2&=&-4 \left(13 m^8-17 (s+u) m^6+\left(6(s^2+u^2)+20 u s\right) m^4-7 s u
(s+u) m^2+3 s^2 u^2\right),\\
a_3&=&-8 \left(m^2-s\right)^2 \left(m^2-u\right)^2,\\
a_4&=&\left(m^2-s\right) \left(m^2-u\right) \left(m^2 (s+u)-2 s u\right).\label{ca4}
\label{Gl2}
\end{eqnarray}
The expression \eqref{m2nkr12g} coincides with the one previously reported
in \cite{DelgadoAcosta:2010nx} where the  spin-$1/2$ particles in the
$(1/2,0)\oplus(0,1/2)$ representation space have been allowed to be of an arbitrary $g$-factor.
%Also in  \cite{DelgadoAcosta:2010nx}, the $g$ value has been fixed to $g=2$ from the requirement on
%the asymptotic vanishing of the Compton scattering amplitudes in the ultraviolet.
The result of the current section, in combination with the causality proof of the
relevant wave equation delivered above,
completes the consistency proof of the combined- Lorentz-and Poincar\'e invariant projector
method in its application to the $(1/2,0)\oplus(0,1/2)$ sector of the four-vector spinor.
%%%%%%%%%%%%%%%%%%%%%%%%%%%%%%%%%%%%%%%%%%%%%%%%%%%%%%%%%%%%%%%%%%%%%%%%%%%%%%%%%%%%%%%%%%%%%%%%%%%%%%%%%%%%%%%%%%%%%%%
%%%%%%%%%%%%%%%%%%%%%%%%%%%%%%%%%%%%%%%%%%%%%%%%%%%%%%%%%%%%%%%%%%%%%%%%%%%%%%%%%%%%%%%%%%%%%%%%%%%%%%%%%%%%%%%%%%%%%%%
\subsection{Compton scattering off spin-$1/2$ particles in the two-spin valued
 { $(1/2,1)\oplus(1,1/2)$} {}  sector of the four-vector--spinor}\label{sec5b}
\begin{figure}[ht]
\includegraphics{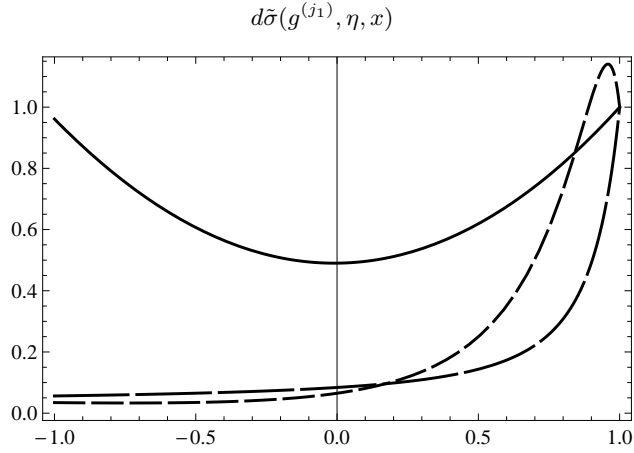}
\caption{\label{dsfig} Differential cross section for particles in the
$(1/2,j_1)\oplus(j_1,1/2)$ sector of the four-vector spinor as a function of
$x=\cos\q$  (where $\q$ is the scattering angle). The solid curve represents
the classical limit, this is, the differential cross section
$d\widetilde{\s}(g^{(j_1)},\h,x)$ from \eqref{dst} at $\h=\omega/m=0$ (where
$\omega$ is the energy of the incident photon), the long-dashed curve
corresponds to $d\widetilde{\s}(g^{(0)},\h,x)$ with $g^{(0)}=2$ at the energy of
$\h=4$, and the short-dashed curve represents $d\widetilde{\s}(g^{(1)},\h,x)$
with $g^{(1)}=-2/3$ also at the energy of $\h=4$. The differential cross section 
has the correct Thompson limit for any $g^{(j_1)}$ value.}
\end{figure}

\begin{figure}[ht]
\includegraphics{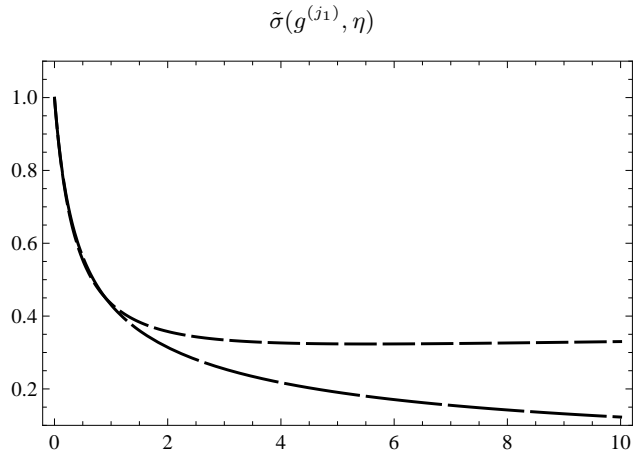}
\caption{\label{sfig} Total cross section $\widetilde{\s}(g^{(j_1)},\h)$
\eqref{tcst} for particles in the
$(1/2,j_1)\oplus(j_1,1/2)$ sector of the four-vector spinor as a function of
$\h=\omega/m$
(where $\omega$ is the energy of the incident photon) up to $\h=10$. The
long-dashed curve correspond
to $\widetilde{\s}(g^{(0)},\h)$ with $g^{(0)}=2$, and the short-dashed curve
represents
$\widetilde{\s}(g^{(1)},\h)$ with $g^{(1)}=-2/3$. While $\widetilde{\s}(g^{(0)},\h)$ is vanishing in the 
ultrarelativistic limit, $\widetilde{\s}(g^{(1)},\h)$ approaches a fixed value, and compatible with unitarity.
}
\end{figure}

%%%%%%%%%%%%%%%%%%%%%%%%%%%%%%%%%%%%%%%%%%%%%%%%%%%%%%%%%%%%%%%%%%%%%%%%%%%%%%%%%%%%%%%%%%%%%%%%%%%%%%%%%%%%%%%%%%%%%%%
The averaged squared amplitude in this case is elaborated applying the method
already presented in the previous section. Using again \eqref{m2av}-\eqref{phpr}, now for
$j_1=1$, gives,
\begin{eqnarray}\label{m2nkr12j1}
\overline{\vert{\mathcal M}^{(1)}(g^{(1)})\vert^2}&=& f_0+f_D
+\frac{e^4(2 m^2-s-u)}{16 m^2 \left(m^2-s\right)^2
\left(m^2-u\right)^2}\sum_{k=1}^4 (g^{(1)}-2)^k a_{k},
\end{eqnarray}
with $g^{(1)}=-2/3$, same as before in \eqref{g1}, and with the same $a_k$ coefficients in \eqref{ca1}-\eqref{ca4}. The above expression
coincides in form with \eqref{m2nkr12g} and with the result previously reported
in \cite{DelgadoAcosta:2010nx}, in the particular case of $g=-2/3$. Obtaining the differential
cross-section in the laboratory frame from squared amplitudes of the types in \eqref{m2nkr12g} and
\eqref{m2nkr12j1} is  straightforward (see \cite{DelgadoAcosta:2010nx} for details).
After some algebraic manipulations one arrives at,
\aeq\label{dss}
\frac{d \s(g^{(j_1)},\h,x)}{d \Omega}=z_0+z_D+\frac{(x-1) r_0^2}{64 ((x-1) \eta
-1)^3}\sum_{k=1}^4 (g^{(j_1)}-2)^k b_k,
\ceq
where $r_0=e^2/(4\p m)=\a m$, $\h=\o/m$ where $\o$ is the energy if the incident 
photon and $x=\cos\q$, being $\q$ the scattering angle in the laboratory frame. 
In \eqref{dss} $z_0$ denotes the standard differential cross-section for Compton scattering
off spin-0 particles and $(z_0+z_D)$  is the standard differential cross-section for
Compton scattering off Dirac particles, this is 
\begin{eqnarray}
z_0&=&\frac{\left(x^2+1\right) r_0^2}{2 ((x-1) \eta -1)^2},\label{z0}\\
z_D&=&-\frac{(x-1)^2 \eta ^2 r_0^2}{2 ((x-1) \eta -1)^3}. \label{zd}
\end{eqnarray}
We further have introduced the following notations,
\begin{align}
b_1=&-32 (x-1) \eta ^2,\\
b_2=&4 \left(x^2-3 x+8\right) \eta ^2,\\
b_3=&16 \eta ^2,\\
b_4=&(x+3) \eta ^2.
\end{align}
The differential cross-section \eqref{dss} has the following properties:
\aec
\lim_{x\rightarrow 1} \frac{d \s(g^{(j_1)},\h,x)}{d \Omega}&=&r_0^2,\\
\lim_{\h\rightarrow 0}\frac{d \s(g^{(j_1)},\h,x)}{d
\Omega}&=&\frac{r_0^2}{2}(x^2+1),\\
\lim_{\h\rightarrow \infty}\frac{d \s(g^{(j_1)},\h,x)}{d \Omega}&=&0,
\cec
meaning that in the forward direction ($x=\cos\q =1$) it takes the  $r_0^2$ value.
In the classical $\h \rightarrow 0$ limit the differential cross section is symmetric with respect to
the scattering angle $\q$, while  in the high energy $\h\rightarrow
\infty$ limit it vanishes independently of the $g^{(j_1)}$ factor value.
This observation applies to each one of the two $j_1=0$-, and $j_1=1$ sectors of $\psi_\mu$
considered here,  and the related  $g^{(0)}=2$ and $g^{(1)}=-2/3$ values.
The behavior of the differential
cross-section is displayed  in  Fig. \ref{dsfig}, which is a plot of,
\aeq
d\widetilde{\s}^{(j_1)}\equiv\frac{1}{r_0^2}\frac{d \s(g^{(j_1)},\h,x)}{d
\Omega}, \quad j_1=0,1\label{dst}.
\ceq
and $g^{(0)}=2$, $g^{(1)}=-2/3$.
%^^^^^^^^^^^^^^^^^^^^^^^^^^^^^^^^^^^^^^^^^^^^^^^^^^^^^^^^^^^^^^^^^^^^^^^^^^^^^^^^^^^^^^^^^^^^^^^^^^
%----------------------Figure----------------------------------------------------------------------
%----------------------Figure----------------------------------------------------------------------
%^^^^^^^^^^^^^^^^^^^^^^^^^^^^^^^^^^^^^^^^^^^^^^^^^^^^^^^^^^^^^^^^^^^^^^^^^^^^^^^^^^^^^^^^^^^^^^^^^^
Integration of \eqref{dss} over the solid angle leads to the total cross-sections,
\aeq\label{tcs}
\s(g^{(j_1)},\h)=s_0+s_D
+\sum_{k=1}^4 (g^{(j_1)}-2)^k\(\frac{c_k}{128 \eta  (2 \eta +1)^2}
+\frac{\log (2 \eta +1)h_k}{256 \eta ^2}\)3 \s_T,
\ceq
where $\s_T$ stands for the Thompson cross section $\s_T=(8/3)\pi r_0^2$.
The following notations have been used,
\aec
s_0&=&\frac{3 (\eta +1) \s_T (2 \eta  (\eta +1)
-(2 \eta +1) \log (2 \eta +1))}{4 \eta ^3 (2 \eta +1)},\label{s0}\\
s_D&=&\frac{3 \s_T \left((2 \eta +1)^2 \log (2 \eta +1)
-2 \eta  (3 \eta +1)\right)}{8 \eta  (2 \eta +1)^2}\label{sd},
\cec
where $s_0$ and $(s_0+s_D)$ are the standard cross-sections for Compton
scattering off spin-$0$ and spin-$1/2$ Dirac particles, while
the $c$ and $h$ coefficients stand for the following quantities,
\aec
c_1&=&-32 \eta  (3 \eta +1),\\
c_2&=&4 \left(6 \eta ^3+\eta ^2+8 \eta +3\right),\\
c_3&=&16 \eta ^3,\\
c_4&=&\eta  \left(4 \eta ^2+3 \eta +1\right),\\
h_1&=&32 \eta,\\
h_2&=&4 (\eta -3),\\
h_3&=&0,\\
h_4&=&-\eta.
\cec
The total cross section \eqref{tcs} has the following limits,
\aec
\lim_{\h\rightarrow 0}\s(g^{(j_1)},\h)&=&\s_T,\\
\lim_{\h\rightarrow
\infty}\s(g^{(j_1)},\h)&=&\frac{3}{128}(g^{(j_1)}-2)^2((g^{(j_1)})^2+2)\s_T.
\cec
Consequently, while in the $g^{(j_1)}=2$ case the cross section was vanishing, for
$g^{(j_1)}=-2/3$, it approaches  $\frac{11\s_T}{27}$. In
Fig.~\ref{sfig} the following quantity is plotted,
\aeq\label{tcst}
\widetilde{\s}(g^{(j_1)},\h)\equiv \frac{1}{\s_T}\s(g^{(j_1)},\h).
\ceq
For  $g^{(j_1)}=2$ one observes the usual decreasing behavior of the Dirac cross section
with energy increase,
while for $g^{(j_1)}=-2/3$ i.e. for
spin-$1/2$ in   $(1/2,1)\oplus(1,1/2)$, the cross section $\widetilde{\s}(g^{(j_1)},\h)$ at high energy
approaches the fixed value of $\frac{11}{27}$ as one can see in the Fig. \ref{sfig}.
%^^^^^^^^^^^^^^^^^^^^^^^^^^^^^^^^^^^^^^^^^^^^^^^^^^^^^^^^^^^^^^^^^^^^^^^^^^^^^^^^^^^^^^^^^^^^^^^^^^
%----------------------Figure----------------------------------------------------------------------
%----------------------Figure----------------------------------------------------------------------
%^^^^^^^^^^^^^^^^^^^^^^^^^^^^^^^^^^^^^^^^^^^^^^^^^^^^^^^^^^^^^^^^^^^^^^^^^^^^^^^^^^^^^^^^^^^^^^^^^^
%%%%%%%%%%%%%%%%%%%%%%%%%%%%%%%%%%%%%%%%%%%%%%%%%%%%%%%%%%%%%%%%%%%%%%%%%%%%%%%%%%%%%%%%%%%%%%%%%%%%%%%%%%%%%%%%%%%%%%%
\subsubsection{Allowing for an arbitrary  {$g$}-factor for
spin-$1/2$ in  { $(1/2,1)\oplus(1,1/2)$} {}}\label{sec5b1}
%%%%%%%%%%%%%%%%%%%%%%%%%%%%%%%%%%%%%%%%%%%%%%%%%%%%%%%%%%%%%%%%%%%%%%%%%%%%%%%%%%%%%%%%%%%%%%%%%%%%%%%%%%%%%%%%%%%%%%%
Testing causality of the wave equation for spin-$1/2$ in  $(1/2,1)\oplus(1,1/2)$  was
possible not only because of the specifically chosen antisymmetric part of the
$[\G^{(1)}_{\m\n}]_{\a\b}$ tensor, but also because of the
auxiliary condition \eqref{s1aux}, which takes its origin from the relation
\aeq
\g^\a[\G^{(1)}_{\m\n}]_{\a\b}=0.\label{gaux}
\ceq
In being momentum independent, \eqref{gaux} ensures that
the particle always belongs to  $(1/2,1)\oplus(1,1/2)$. There is a
number of antisymmetric structures available in the four-vector spinor representation space (see for
example \cite{DelgadoAcosta:2009ic} for details)
which we can employ  to build up an extension
of the $[\G^{(1)}_{\m\n}]_{\a\b}$ tensor toward an arbitrary $g$ value.
However, the majority of  these extensions do not allow us to perform
the factorization \eqref{j1cfact} (essential for providing the causality proof of the propagation
within an electromagnetic field) nor would they  satisfy \eqref{gaux}. In fact
there is only one acceptable tensor for this purpose and it reads,
\aeq
-\g_\a\g_\m \g_\n \g_\b +2\g_\n \g_\b g_{\a\m}-2\g_\m\g_\b g_{\a\n}+\g_\a
\g_\b g_{\m\n}\label{ext1}.
\ceq
Notice however that, by virtue of
\aeq\label{f1ort}
\g^\a [f^{(1)}(p)]_\a=0,
\ceq
an extension based on  \eqref{ext1}  does not provide any
contribution neither to the on-shell current of the type of \eqref{j1ov},
nor to the related magnetic dipole moment. A similar situation is observed
with regard to  the Compton scattering amplitude
\eqref{camp} where also there,  all contributions coming from the
extension are vanishing, ultimately because of \eqref{f1ort}.
{}From this  we conclude that any  extension toward an arbitrary $g$-factor
that is compatible with our causality testing procedure, is irrelevant to both the
magnetic dipole moment-- , and  the  Compton scattering cross section values.
For this reason we restrict  ourselves
to the calculation of Compton scattering off spin-$1/2$ in  $(1/2,1)\oplus(1,1/2)$
presented above in reference to the  $g^{(1)}=-2/3$ value following  from  \eqref{g1}.

%%%%%%%%%%%%%%%%%%%%%%%%%%%%%%%%%%%%%%%%%%%%%%%%%%%%%%%%%%%%%%%%%%%%%%%%%%%%%%%%%%%%%%%%%%%%%%%%%%%%%%%%%%%%%%%%%%%%%%%
%%%%%%%%%%%%%%%%%%%%%%%%%%%%%%%%%%%%%%%%%%%%%%%%%%%%%%%%%%%%%%%%%%%%%%%%%%%%%%%%%%%%%%%%%%%%%%%%%%%%%%%%%%%%%%%%%%%%%%%
%%%%%%%%%%%%%%%%%%%%%%%%%%%%%%%%%%%%%%%%%%%%%%%%%%%%%%%%%%%%%%%%%%%%%%%%%%%%%%%%%%%%%%%%%%%%%%%%%%%%%%%%%%%%%%%%%%%%%%%
\section{Conclusions}\label{sec6}
%%%%%%%%%%%%%%%%%%%%%%%%%%%%%%%%%%%%%%%%%%%%%%%%%%%%%%%%%%%%%%%%%%%%%%%%%%%%%%%%%%%%%%%%%%%%%%%%%%%%%%%%%%%%%%%%%%%%%%%
In the present work we studied the status of the lower-spin components  of the Rarita-Schwinger
four-vector spinor $\psi_\mu$, a Lorentz-invariant representation space reducible according
to \eqref{RS_red}. The three criteria we proposed  to qualify representations of
the Lorentz algebra
for the description of physical and observable particles of spin-$s$ are:

\begin{itemize}

\item[(i)] Irreducibility,

\item[(ii)] Hyperbolicity and causality of the related wave equations,

\item[(iii)] Finiteness of the  Compton scattering cross sections in all directions
and in the ultra relativistic limit.

\end{itemize}

In order to fulfill the first criterion we extended the method of the Poincar\'e covariant
spin-$s$ and mass-$m$  projectors \cite{Napsuciale:2006wr} to include momentum independent Lorentz-invariant projectors,
i.e. projectors constructed from the parity conserving Casimir invariant of the Lorentz-algebra.
In so doing we found two quadratic in the momenta wave equations for the
two spin-1/2 sectors in $\psi_\mu$, bi-linearized by properly constructed $4\times 4$ matrices in \eqref{fequations},
the first associated with the single-spin Dirac representation space,
$(1/2,0)\oplus (0,1/2)$, and the second for the spin-$1/2$ companion to spin-$3/2$ in
$(1/2,1)\oplus (1,1/2)$. We demonstrated  hyperbolicity and causality of both equations.
We then showed that the electromagnetic current and the Compton scattering amplitudes of the
first lower spin coincide with those of a genuine Dirac particle, and are characterized by a $g=2$ value, as it should be, and concluded on its
observability. Finally we calculated Compton scattering off the second spin-1/2 in $\psi_\mu$, and
for a gyromagnetic ratio of $g=-2/3$ could find finite
cross sections in all directions and the ultraviolet limit.
Therefore, the observability of the latter state is not excluded by none of the above three criteria.
As long as the spin-$1/2$ under discussion is the companion to
the observable spin-$3/2$ of equal rights
within the irreducible representation space
$(1/2,1)\oplus (1,1/2)$ (the two states are related by ladder
operators), we conclude that all its properties strongly point towards its physical nature.

We furthermore notice that the method of the combined Lorentz-and Poincar\'e invariant projectors is
suitable for the description of fermions of any spin by quadratic equations for sufficiently large
Lorentz algebra representations equipped by  separate Lorentz- and Dirac spinor indices.
 For example, pure spin-$3/2$ can be embedded into the totally antisymmetric tensor of second rank with Dirac spinor components,
$\Psi_{\left[ \mu\nu\right] }$, a representation space that is reducible according to
\begin{eqnarray}
\Psi_{\left[ \mu\nu\right] }&\sim&
\left[ (1,0)\oplus (0,1)\right]\otimes \left[\left(\frac{1}{2},0\right)\oplus \left(0, \frac{1}{2} \right) \right]\nonumber\\
&\longrightarrow &
\left[\left(\frac{1}{2},0\right)\oplus \left(0, \frac{1}{2} \right) \right]\oplus
\left[\left( 1,\frac{1}{2}\right)\oplus \left(\frac{1}{2}, 1 \right) \right]
\oplus \left[\left(\frac{3}{2},0\right)\oplus \left(0, \frac{3}{2} \right) \right].
\label{tensor_spinor}
\end{eqnarray}
The two redundant irreducible subspaces accompanying the single spin-$3/2$ in $\Psi_{\left[ \mu\nu\right]}$ can be 
projected out by momentum independent Lorentz-invariant projectors constructed along the lines of section \ref{sec2b}, 
while the $(3/2,0)\oplus (0,3/2)$ subspace can be identified by the Poincar\'e covariant projector which is second 
order in the momenta. Similarly, spin-$5/2$ can be embedded in the totally antisymmetric Lorentz tensor of second rank 
with four-vector-spinor components, $\Psi_{\left[\mu\nu \right]\eta}$, a representation space reducible according to
\begin{eqnarray}
\Psi_{\left[ \mu\nu\right]\eta }&\sim&
\left[ (1,0)\oplus (0,1)\right]\otimes
\left[ \left(\frac{1}{2},\frac{1}{2} \right)\otimes
\left[\left(\frac{1}{2},0\right)\oplus \left(0, \frac{1}{2} \right) \right]\right]\nonumber\\
&\longrightarrow &
2\left[\left(\frac{1}{2},0\right)\oplus \left(0, \frac{1}{2} \right) \right]\oplus
3\left[\left( 1,\frac{1}{2}\right)\oplus \left(\frac{1}{2}, 1 \right) \right]\nonumber\\
&\oplus& \left[\left(1,\frac{3}{2}\right)\oplus \left(\frac{3}{2}, 1 \right) \right]
\oplus \left[\left(2,\frac{1}{2}\right)\oplus \left(\frac{1}{2}, 2 \right) \right].
\label{tensor_4Vspinor}
\end{eqnarray}
In this case, all the invariant subspaces beyond the two-spin sector $(2,1/2)\oplus (1/2,2)$
can be projected out by properly constructed momentum independent Lorentz-invariant
projectors while spin-$3/2$, and spin-$5/2$ in  $(2,1/2)\oplus (1/2,2)$ can be separated by
means of a Poincar\'e covariant projector. In this fashion, a second order Lagrangian
for spin-$5/2$ description in terms of a representation space of separate Lorentz and Dirac 
spinor indices can be furnished. 

We expect relevance of our observations in processes which are sensitive to the irreducibility of the Lorentz representations.
%%%%%%%%%%%%%%%%%%%%%%%%%%%%%%%%%%%%%%%%%%%%%%%%%%%%%%%%%%%%%%%%%%%%%%%%%%%%%%%%%%%%%%%%%%%%%%%%%%%%%%%%%%%%%%%%%%%%%%%
%%%%%%%%%%%%%%%%%%%%%%%%%%%%%%%%%%%%%%%%%%%%%%%%%%%%%%%%%%%%%%%%%%%%%%%%%%%%%%%%%%%%%%%%%%%%%%%%%%%%%%%%%%%%%%%%%%%%%%%
%%%%%%%%%%%%%%%%%%%%%%%%%%%%%%%%%%%%%%%%%%%%%%%%%%%%%%%%%%%%%%%%%%%%%%%%%%%%%%%%%%%%%%%%%%%%%%%%%%%%%%%%%%%%%%%%%%%%%%%

\end{document}